\documentclass[useAMS,usenatbib,usegraphicx]{mn2e}
\usepackage{psfig}   
\usepackage{graphicx}
\usepackage{amssymb}
\usepackage{subfigure}

\title[Planets in substructured clusters]{The effects of dynamical interactions on planets in young substructured star clusters}

\author[R. J. Parker and S. P. Quanz]{Richard J.~Parker\thanks{E-mail: rparker@phys.ethz.ch} and Sascha P. Quanz \vspace*{0.1cm}\\
   Institute for Astronomy, ETH Z{\"u}rich, Wolfgang-Pauli-Strasse 27, 8093 Z{\"u}rich, Switzerland}

\begin{document}

\date{Accepted for publication in MNRAS}
                             
\pagerange{\pageref{firstpage}--\pageref{lastpage}} \pubyear{2011}

\maketitle

\label{firstpage}

\begin{abstract}
We present $N$-body simulations of young substructured star clusters undergoing various dynamical evolutionary scenarios and examine the direct effects 
of interactions in the cluster on planetary systems. We model clusters initially in cool collapse, in virial equilibrium 
and expanding, and place a 1 Jupiter-mass planet at either 5\,au or 30\,au from their host stars, with zero eccentricity. We find that after 10\,Myr $\sim$10 per~cent of planets 
initially orbiting at 30 au have been liberated from their parent star and form a population of free-floating planets. A small number of these planets are captured 
by other stars.  A further $\sim$10 per cent have their orbital eccentricity (and less often their semi-major axis) significantly altered. For planets originally at 5 au the fractions are a factor of 2 lower. 
The change in eccentricity is often accompanied by a change in orbital inclination which may lead to additional dynamical perturbations in planetary systems with multiple planets. The fraction of liberated and 
disrupted planetary systems is highest for subvirial clusters, but virial and supervirial clusters also dynamically process planetary systems, due to interactions in the substructure.  

Of the planets that become free-floating, those that remain observationally associated with the cluster (i.e.\,\,within two half-mass radii of the cluster centre) have a similar velocity distribution to the entire star cluster, irrespective of 
whether they were on a 5\,au or a 30\,au orbit, with median velocities typically $\sim$1\,km\,s$^{-1}$. Conversely, those planets that are no longer associated with the cluster have similar  velocities to the non-associated stars if they were originally at 5\,au 
($\sim$9\,km\,s$^{-1}$), whereas the planets originally at 30\,au have much lower velocities (3.8\,km\,s$^{-1}$) than the non-associated stars (10.8\,km\,s$^{-1}$). These findings highlight potential pitfalls of concluding that 
(a)  planets with similar velocities to the cluster stars represent the very low-mass end of the IMF, and (b) planets on the periphery of a cluster with very different observed velocities form through different mechanisms.
\end{abstract}


\section{Introduction}

A large proportion of stars form in clustered environments \citep[e.g.][]{Lada03,Lada10}. Whether such star-forming regions are dense enough to undergo significant dynamical processing is 
currently under debate \citep{Bressert10}, but dynamical considerations do suggest that some clusters experience a dense phase during their evolution 
\citep[e.g.][]{Kroupa95a,Kroupa99,Moraux07,Allison09b,Parker09a}. Recently, \citet{Allison09b} showed that the mass segregation in the Orion Nebula Cluster (ONC) can be explained if the 
cluster was originally subvirial, and substructured. This causes the cluster to collapse in the first 1\,Myr which leads to dynamical mass segregation and also heavily processes the primordial binary population \citep*{Parker11c}.  

Such a dense phase would have serious implications for the survivability of planetary systems: planets could either be directly liberated from their host stars after a close encounter  
with another star, or the orbital elements of the planets could be significantly altered leaving the planetary system dynamically unstable 
\citep[e.g.][]{Holman97,Innanen97,Takeda05,Malmberg07a,Fabrycky07,Naoz11}.

In terms of density, the most extreme galactic stellar environments are Globular clusters, 
and $N$-body calculations have demonstrated that planets orbiting at distances of between 0.5 and 50\,au would be liberated from their host stars in these clusters \citep{Hurley02}. On the other hand, 
open clusters are not as dense but, as discussed above, may undergo a dense phase in which for example, the primordial stellar binary population can be significantly processed \citep{Kroupa99,Parker11c}. Several authors 
have studied the effects of open cluster environments on planetary systems \citep[e.g.][and references therein]{Laughlin98,Bonnell01b,Smith01,Davies01,Adams06,Fregeau06,Spurzem09}. In the main, these authors 
determined the cross-sectional probability for scattering planets, based on the likelihood of an interaction with a passing single or binary star. 

A small number of authors performed direct $N$-body simulations of clusters containing planets \citep{Hurley02,Spurzem09}, but mainly considered Globular clusters. Almost all these papers assume that the cluster 
is in virial equilibrium and has a smooth radial profile (e.g. a Plummer sphere \citep{Plummer11}, or King profile, \citep{King66}) at birth. \citet{Adams06} considered subvirial clusters, but with smooth 
Plummer spheres, rather than substructured environments. Observations of young star 
 forming regions show them to be highly substructured \citep[e.g.][]{Cartwright04,Schmeja11} and often subvirial \citep[e.g.][]{Peretto06,Proszkow09}. 
 
 In this paper we focus on the dynamical evolution of substructured open clusters, with subvirial, virial and supervirial initial 
 conditions. We perform $N$-body simulations of such clusters and include planets directly in the models. We outline our method for setting up the clusters in Section~\ref{method}, we present 
 our results in Section~\ref{results} and we discuss them in Section~\ref{discuss}, placing them in the context of searches for free-floating planets and planets in open clusters and the field. We draw our 
 conclusions in Section~\ref{conclude}.

\section{Method}
\label{method}

In this section we describe our method of setting up star clusters with substructure, and the assignment of companions (either stellar or planetary) to each star.

\subsection{Cluster set-up}

Observations of young star forming regions indicate that a high level of substructure is present \citep{Cartwright04,Sanchez09,Schmeja11}. A convenient way of creating substructure on all scales is the 
fractal distribution \citep{Goodwin04a}, which makes each location in the cluster statistically identical to any other. Note that we are not claiming that the best 
approximation of substructure in star clusters is the fractal method; rather that the fractal is the most convenient method of producing substructure. Its main advantage is that the substructure 
is defined by just one number; the fractal dimension $D$. This defines how `fractal' the cluster is, with $D = 1.6$ describing a highly clumpy cluster 
(in three dimensions) and $D = 3.0$ describing a uniform sphere.

We set up the fractals according to the method in \citet{Goodwin04a}. This begins by defining a cube of side $N_{\rm div}$ (we adopt $N_{\rm div} = 2.0$ 
throughout), inside of which the fractal is built. A first-generation parent is placed at the centre of the cube, which then spawns $N_{\rm div}^3$ sub-cubes, each containing a first generation 
child at its centre. The fractal is then built by determining which of the children will themselves become parents, and spawn their own offspring. This is determined by the 
fractal dimension, $D$, where the probability that the child becomes a parent is given by $N_{\rm div}^{(D- 3)}$. For a lower fractal dimension fewer children 
will mature and the final distribution will contain more substructure. The mean number of maturing children is $2^D$, and so the preferred values of $D$  are 1.6, 2.0, 2.6 or 3.0, which would 
correspond to the number of maturing children being an integer, and therefore result in fewer departures from the specified fractal dimension \citep{Goodwin04a}.  Any children that do not become parents in a given step are removed, along with all of their parents. A small amount of noise is then 
added to the positions of the remaining children, preventing the cluster from having a gridded appearance and the children become parents of the next generation. Each new parent 
then spawns $N_{\rm div}^3$ second-generation children in $N_{\rm div}^3$ sub-subcubes, with each second-generation child having a $N_{\rm div}^{(D - 3)}$ 
probability of becoming a second generation parent. This process is repeated until there are substantially more children than are required. The children are pruned to produce a 
sphere from the cube and are then randomly removed (so maintaining the fractal dimension) until the required number of children is left. These children then become stars in the 
cluster. 

We set up clusters with just one fractal dimension, $D = 2.0$, which gives the cluster a moderate level of substructure, but is not as extreme as, for example, a cluster with 
$D = 1.6$. The effect of varying the fractal dimension is to change the level of dynamical interactions that take place as the cluster evolves. \citet{Allison09b,Allison10} 
show that the lower the fractal dimension, the more likely (and quickly) it is that dynamical mass segregation will occur. Trapezium systems can also form dynamically within 1\,Myr in a 
cluster undergoing cool collapse with a low fractal dimension \citep[$D \leq 2.0$,][]{Allison11}. A higher proportion of primordial binary systems are also disrupted if the fractal dimension is lower \citep{Parker11c}. However, we adopt 
a mid-range value of $D = 2.0$ throughout this work.

To determine the velocity structure of the cluster, children inherit their parent's velocity plus a random component that decreases with each generation of the fractal. The children of the 
first generation are given random velocity components from a Gaussian of mean zero. The random component added to the children's velocity is also 
drawn from a Gaussian, but is then multiplied by $1/N_{\rm div}$ for each generation of the fractal. This results in a velocity structure in which nearby stars have similar velocities, but distant stars can have very 
different velocities. The velocity of every star is scaled to obtain the desired virial ratio of the cluster.

We examine three different evolutionary scenarios for our clusters. In the main, we adopt the cool-collapse scenario as advocated for the ONC by \citet{Allison09b,Allison10}.  Such 
 clusters have a virial ratio of $Q = 0.3$, where we define the virial ratio as $Q = T/|\Omega|$ ($T$ and $|\Omega|$ are the total kinetic energy and total potential energy of the 
 stars/planets, respectively). Therefore, a cluster with $Q = 0.5$ is in virial equilibrium and a cluster with $Q = 0.3$ is said to be subvirial, or `cool'. We also set up supervirial (`warm') clusters 
 with a virial ratio of $Q = 0.7$. 

\begin{figure}
 \begin{center}
\rotatebox{270}{\includegraphics[scale=0.39]{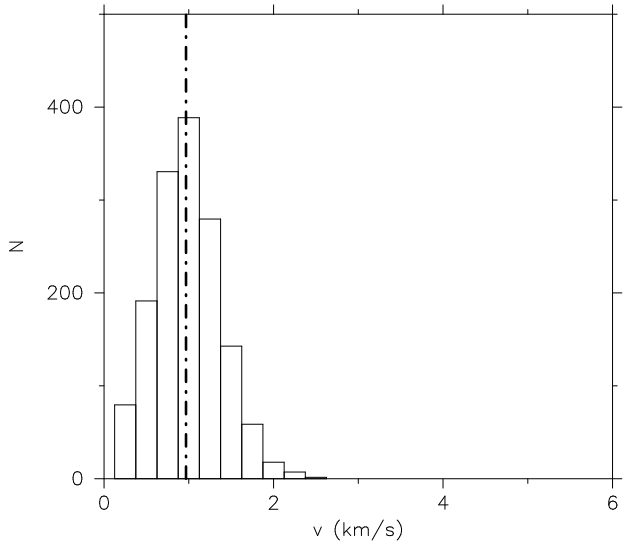}}
\caption[bf]{The initial velocity distribution for our `default' cluster, a subvirial ($Q = 0.3$), $N_{\rm obj} = 1500$ fractal cluster. We have used the binary  
centre-of-mass velocity in the histogram, rather than the individual velocities. We show the median velocity (0.98\,km\,s$^{-1}$) in the cluster by the dot--dashed line.}
\label{cluster_velocities}
\end{center}
\end{figure}

The initial velocity distribution for a subvirial ($Q = 0.3$), $N_{\rm obj} = 1500$ cluster is shown in Fig.~\ref{cluster_velocities}. We bin the centre-of-mass velocities, 
rather than the individual velocities of the binary (either star-star or star-planet) components, as these represent the system velocities. The median cluster velocity (0.98\,km\,s$^{-1}$) is shown by the dot-dashed line. 



\subsection{System properties}

Our `typical' star cluster has a total of 750 primary stars, with masses drawn from a 3--part \citep[e.g.][]{Kroupa02} IMF of the form:
\begin{equation}
 N(M)   \propto  \left\{ \begin{array}{ll} 
 M^{-0.3} \hspace{0.4cm} m_0 < M/{\rm M_\odot} \leq m_1   \,, \\ 
 M^{-1.3} \hspace{0.4cm} m_1 < M/{\rm M_\odot} \leq m_2   \,, \\ 
 M^{-2.3} \hspace{0.4cm} m_2 < M/{\rm M_\odot} \leq m_3   \,,
\end{array} \right.
\end{equation}
and we choose $m_0$ = 0.08\,M$_\odot$, $m_1$ = 0.1\,M$_\odot$, $m_2$ = 0.5\,M$_\odot$, and  $m_3$ =
50\,M$_\odot$. We adopt an upper limit of 50\,M$_\odot$ because our default model is an Orion-like cluster, where the most massive star, $\theta^1$\,Ori\,C, has a system mass of 45--50\,M$_\odot$ \citep{Kraus07,Kraus09}\footnote{$\theta^1$\,Ori\,C 
is likely to be a triple system, where the most massive component has a mass of 30 -- 35\,M$_\odot$ \citep{Kraus07,Lehmann10}.}. 
Also,  we randomly sample from the IMF, which in principle could lead to the most massive star dominating the mass of the entire cluster if we adopt an upper limit 
of either 150\,M$_\odot$ \citep{Figer05}, or 300\,M$_\odot$ \citep{Crowther10}.  In some simulations, we vary the number of stars, with one suite of clusters containing 1500 primary stars initially, and another containing 100 primary stars. 
This samples two different cluster mass regimes, and combined with our default model of 750 stars gives us three different cluster densities (as all the clusters have a radius of $\sim\,1$\,pc). 

Clusters are observed to have a $-2$ power law mass distribution from 10$^2$ to 10$^5$\,M$_\odot$, so an equal mass of stars make up the Galactic field from all clusters \citep{Lada03}. Our models therefore cover the first three orders of magnitude in cluster mass; 
we do not simulate more massive clusters due to computational limitations.

\subsubsection{Stellar companions}

For each primary star, we assign a companion based on the binary fraction associated with its mass. We divide primaries into four groups, roughly corresponding to the binary fraction of these systems observed in the Galactic 
field\footnote{We note that the binary fraction in most clusters may be much higher, in some cases approaching 100 per cent \citep[][and references therein]{Kroupa08,Goodwin10}. 
We will address the issue of planets orbiting the component(s) of stellar binary systems in star clusters in a future paper.}. 
Primary masses in the range 
0.08~$\leq M/{\rm M}_\odot~<$~0.47 are M-dwarfs, with a binary fraction of 0.42 \citep{Fischer92}. K-dwarfs have masses in  the range  0.47~$\leq~M/{\rm M}_\odot$~$<$~0.84 
and binary fraction of 0.45 \citep{Mayor92}. We combine G-, F- and A-stars together, and so primary stars with masses from 0.84~$\leq~M/{\rm M}_\odot~\leq$~2.5  are assigned a binary 
 fraction of 0.57 \citep{Duquennoy91}. All stars more massive than  2.5\,M$_\odot$ are grouped together and assigned a binary fraction of unity, as massive stars have a much larger 
 binary fraction than low-mass stars \citep[e.g.][and references therein]{Abt90,Mason98,Kouwenhoven05,Kouwenhoven07,Pfalzner07,Mason09}. 
 
 Secondary masses in these stellar binary systems are drawn from a flat mass ratio distribution; recent work by \citet{Reggiani11} has shown the companion mass ratio of binary stars in the field to 
 be consistent with being drawn from a flat distribution, rather than random pairing from the IMF. We note that drawing companions from a flat distribution means we do not recover 
a Kroupa IMF.
 
 In accordance with the observations of \citet{Duquennoy91} and \citet{Raghavan10}, the periods of binary  star systems are drawn from  the following period generating function:
\begin{equation}
f\left({\rm log_{10}}P\right) \propto {\rm exp}\left \{ \frac{-{({\rm log_{10}}P -
\overline{{\rm log_{10}}P})}^2}{2\sigma^2_{{\rm log_{10}}P}}\right \},
\end{equation}
where $\overline{{\rm log_{10}}P} = 4.8$, $\sigma_{{\rm log_{10}}P} = 2.3$ and $P$ is  in days. We convert the periods to semi-major axes using the the masses of the two components (with $\overline{{\rm log_{10}}P} = 4.8$ 
corresponding to a semi-major axis of roughly 30\,au). 

The eccentricities of intermediate- and wide-separation stellar binaries in the field are well approximated by a thermal distribution \citep{Heggie75,Kroupa08}:
\begin{equation}
f_e(e) = 2e.
\end{equation}
In the sample of \citet{Duquennoy91}, short-separation binaries are observed to be on circular orbits, which we account for by reselecting the eccentricity of a system if in exceeds the 
following period-dependent value \citep{Parker09c}:
\begin{equation}
e_{\rm tid} = \frac{1}{2}\left[0.95 + {\rm tanh}\left(0.6{\rm log_{10}}P - 1.7\right)\right].
\end{equation}
 
 \subsubsection{Planetary companions}
 
 Stars which do not have a stellar companion based on the above criteria are then all assigned a planetary-mass companion. These planets are all 1 Jupiter mass (1\,M$_{\rm Jup} =  9.5 \times 10^{-4}$M$_\odot$), 
are placed on circular orbits ($e = 0$) and have the same initial semi-major axes. In certain simulations, the planets are placed at 5\,au, 
corresponding to a Jupiter--like orbit, or 30\,au, corresponding to a Neptune--like orbit. Due to the stellar system constraints above, no planets are placed around stars with masses exceeding 2.5\,M$_\odot$.

 Comparing our model setup to detected exoplanetary systems in terms of gravitational forces between the planet and the host star we note the following: a 1\,M$_{\rm Jup}$ object around a solar mass star at 
5\,au is similar to the directly imaged exoplanet HR8799 e  \citep{Marois10} assuming its current projected separation corresponds to its true semi-major axis. For a 30\,au orbit the gravitational force is 
comparable to the `b' planet of the HR8799 system \citep{Marois08}. The planets c and d are intermediate cases\footnote{These estimates assume an age of 30 Myr for the HR8799 system in which case the 
masses are roughly 7\,M$_{\rm Jup}$ for the planets c, d, and e, and 5\,M$_{\rm Jup}$ for planet b.}. Other directly imaged planets are either much more weakly bound, as is the case for Fomalhaut b 
\citep{Kalas08} and 1RXS J1609 b \citep{Lafreniere10}, or more tightly bound, in the case of $\beta$ Pictoris b \citep{Lagrange10,Quanz10}, compared to our model setup.

 Whilst creating planetary systems with only one planet is a very simplistic model, creating systems with more than one planet would prohibitively increase the run-time of 
 our simulations, and we consider any change in orbital parameters of a Jupiter-mass planet to be indicative of the general effect of encounters in a star cluster on a 
 fully populated planetary system.   
 
 The simulations are run using the \texttt{kira} integrator in the Starlab package \citep[e.g.][]{Zwart99,Zwart01} for 10\,Myr. We do not include the effects of stellar evolution in the 
 simulations. A summary of all the simulations used in the paper is given in Table~\ref{cluster_props}.

\begin{table}
\caption[bf]{A summary of the different cluster properties adopted for the simulations.
The values in the columns are: the number of stars and planets in each cluster ($N_{\rm obj}$), 
the typical mass of this cluster ($M_{\rm cluster}$), the virial ratio (and corresponding virial `state'; `cool' (c), virialised (v) or `warm' (w)), 
and the separations of the planets.}
\begin{center}
\begin{tabular}{|c|c|c|c|}
\hline 
$N_{\rm obj}$ & $M_{\rm cluster}$ & $Q$  & Planet separations \\
\hline
1500 & $\sim 6 \times 10^2$\,M$_\odot$ & 0.3 (c) & 5\,au  \\
1500 & $\sim 6 \times 10^2$\,M$_\odot$ & 0.3 (c) & 30\,au  \\
\hline
1500 & $\sim 6 \times 10^2$\,M$_\odot$ & 0.5 (v) & 30\,au  \\
1500 & $\sim 6 \times 10^2$\,M$_\odot$ & 0.7 (w) & 30\,au  \\
\hline 
200 & $\sim 10^2$\,M$_\odot$ & 0.3 (c) & 30\,au  \\
3000 & $\sim 10^3$\,M$_\odot$ & 0.3 (c) & 30\,au  \\
\hline
\end{tabular}
\end{center}
\label{cluster_props}
\end{table}

\section{Results}
\label{results}

In this section we will focus on our `default model'; a subvirial ($Q = 0.3$), $N = 1500$ object cluster. We will describe the dynamical evolution of this cluster before describing the effects 
of this evolution on planetary systems. We will then outline the differences in the number of affected planetary systems for the different initial conditions.

\subsection{Cluster evolution}

\begin{figure*}
  \begin{center}
\setlength{\subfigcapskip}{10pt}
\hspace*{-0.3cm}
\subfigure[0\,Myr]{\label{morph_0Myr}\rotatebox{270}{\includegraphics[scale=0.27]{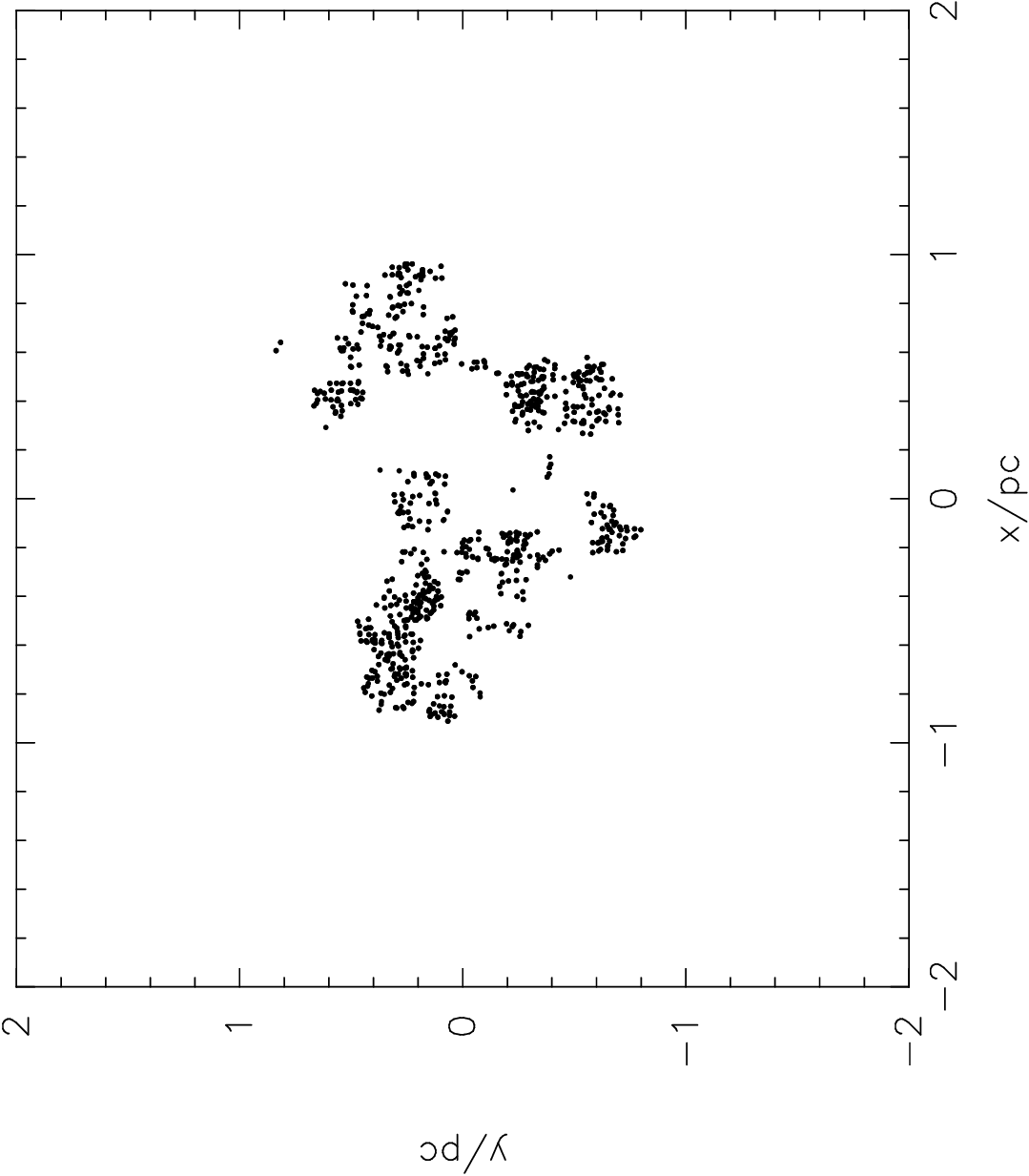}}}
\hspace*{0.2cm}
\subfigure[1\,Myr]{\label{morph_1Myr}\rotatebox{270}{\includegraphics[scale=0.27]{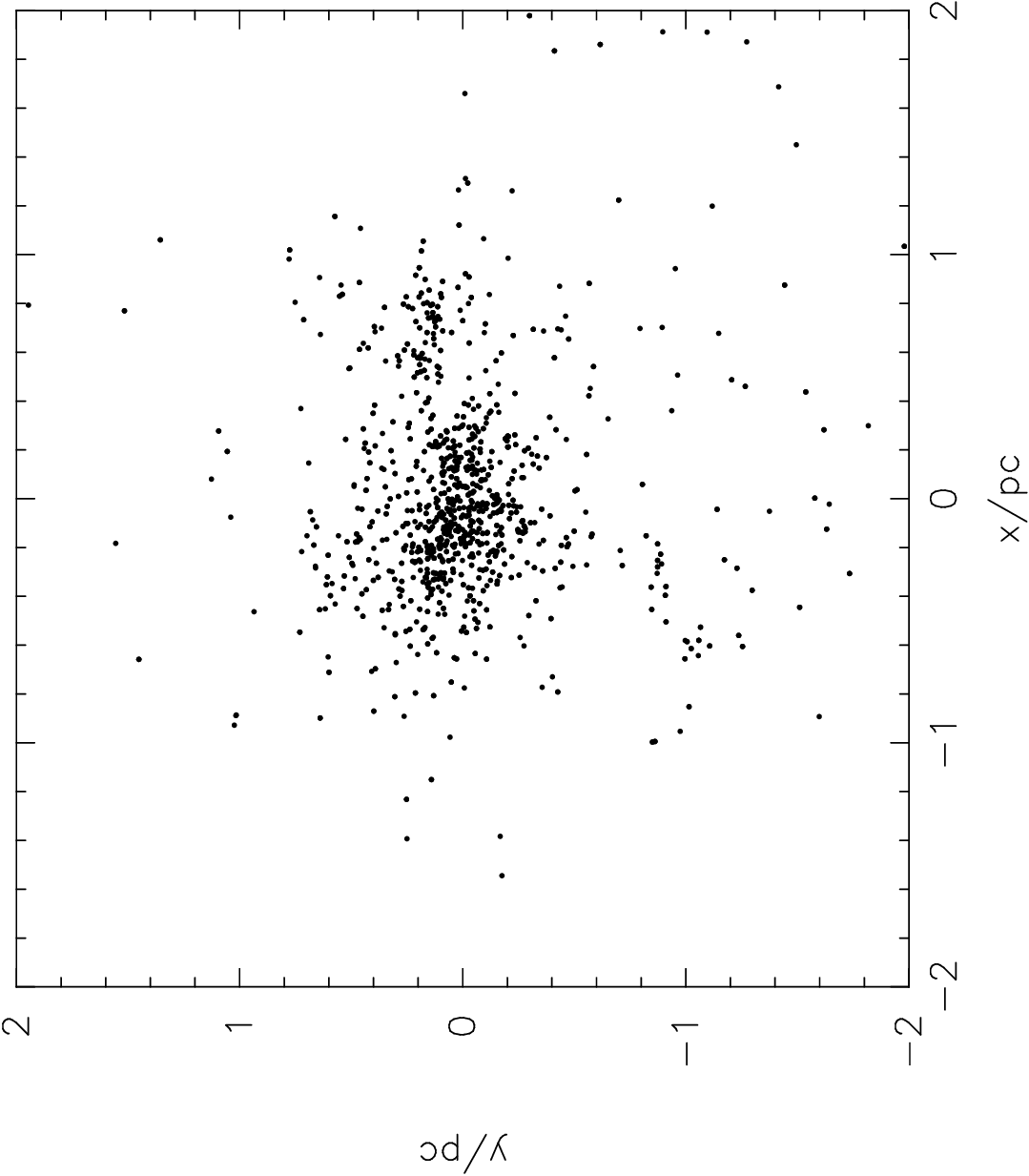}}}
\hspace*{0.2cm}
\subfigure[10\,Myr]{\label{morph_10Myr}\rotatebox{270}{\includegraphics[scale=0.27]{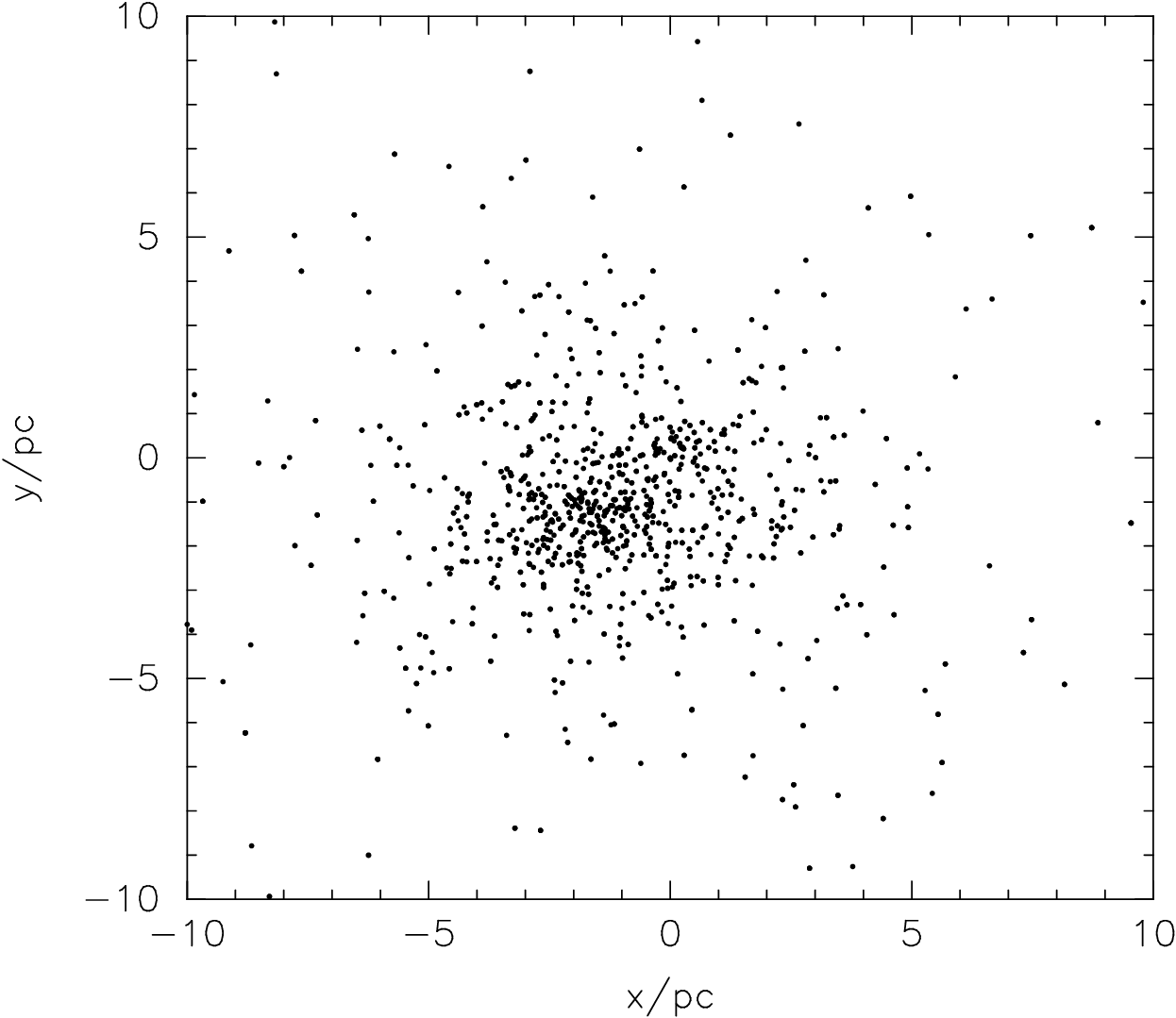}}}
 \end{center}
  \caption[bf]{Typical morphologies for the `default' model ($Q = 0.3$, $N_{\rm obj} = 1500$) clusters at (a) 0\,Myr, (b) 1\,Myr and (c) 10\,Myr. The cluster 
is initially substructured, but this is erased in the first Myr through dynamical interactions in the substructure and the global collapse of the cluster. Note the difference in spatial extent between 1 and 10\,Myr.}
  \label{morph_evolve}
\end{figure*}

In Fig.~\ref{morph_evolve} we show the morphology of a typical cluster undergoing cool collapse. In panel (a) we show the cluster before dynamical evolution, and the substructure is clearly evident. 
After 1\,Myr (panel (b)) the substructure has almost been erased, although the cluster has a similar spatial extent.  However, after 10\,Myr (the time at which we analyse the fraction of affected planetary systems) 
the cluster has expanded following the dense phase of the cool collapse.

\begin{figure*}
 \begin{center}
\setlength{\subfigcapskip}{10pt}
\hspace*{-1.5cm}\subfigure[Half-mass radius]{\label{evolve-a}\rotatebox{270}{\includegraphics[scale=0.39]{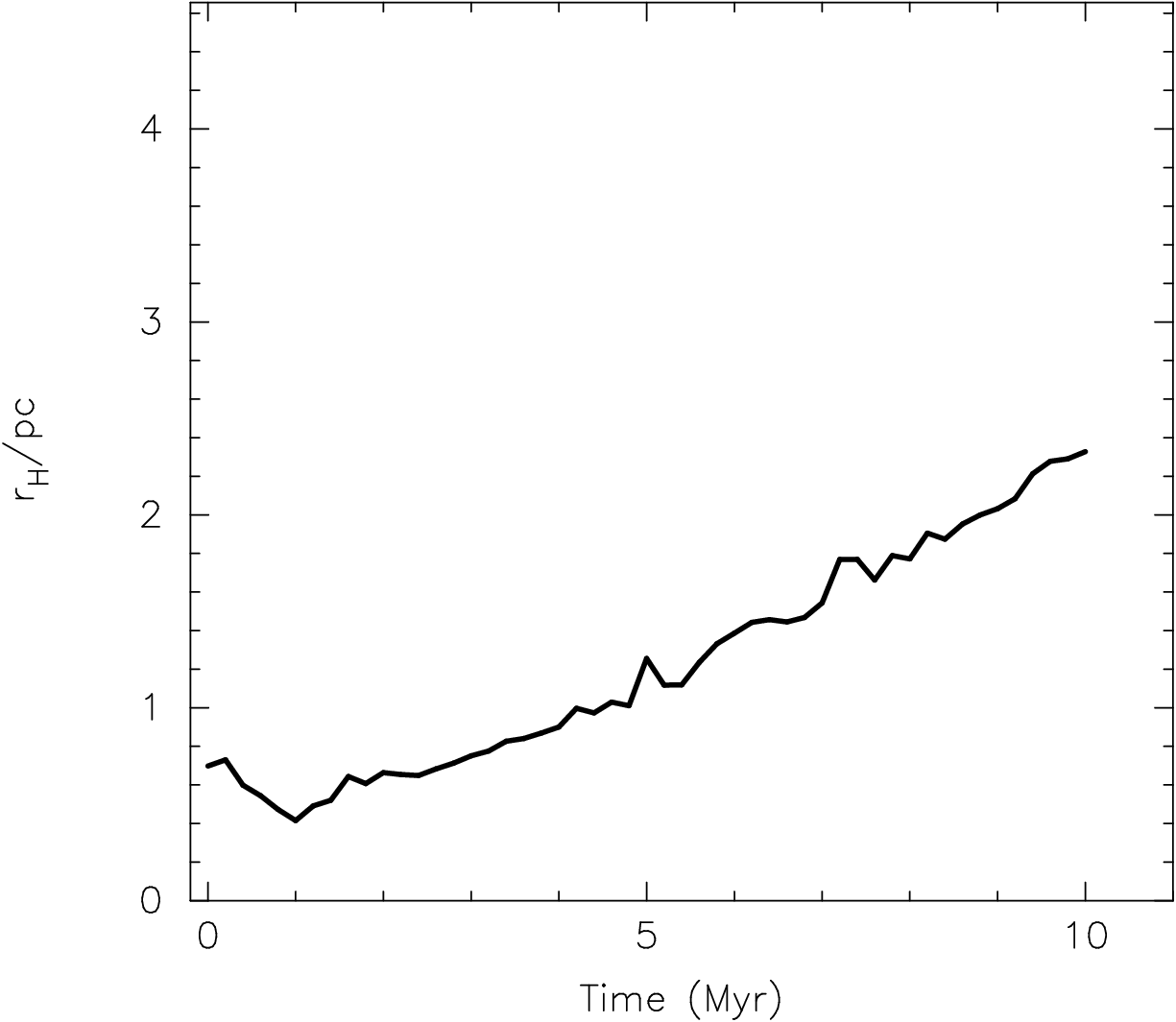}}}
\hspace*{0.3cm} 
\subfigure[Density]{\label{evolve-b}\rotatebox{270}{\includegraphics[scale=0.39]{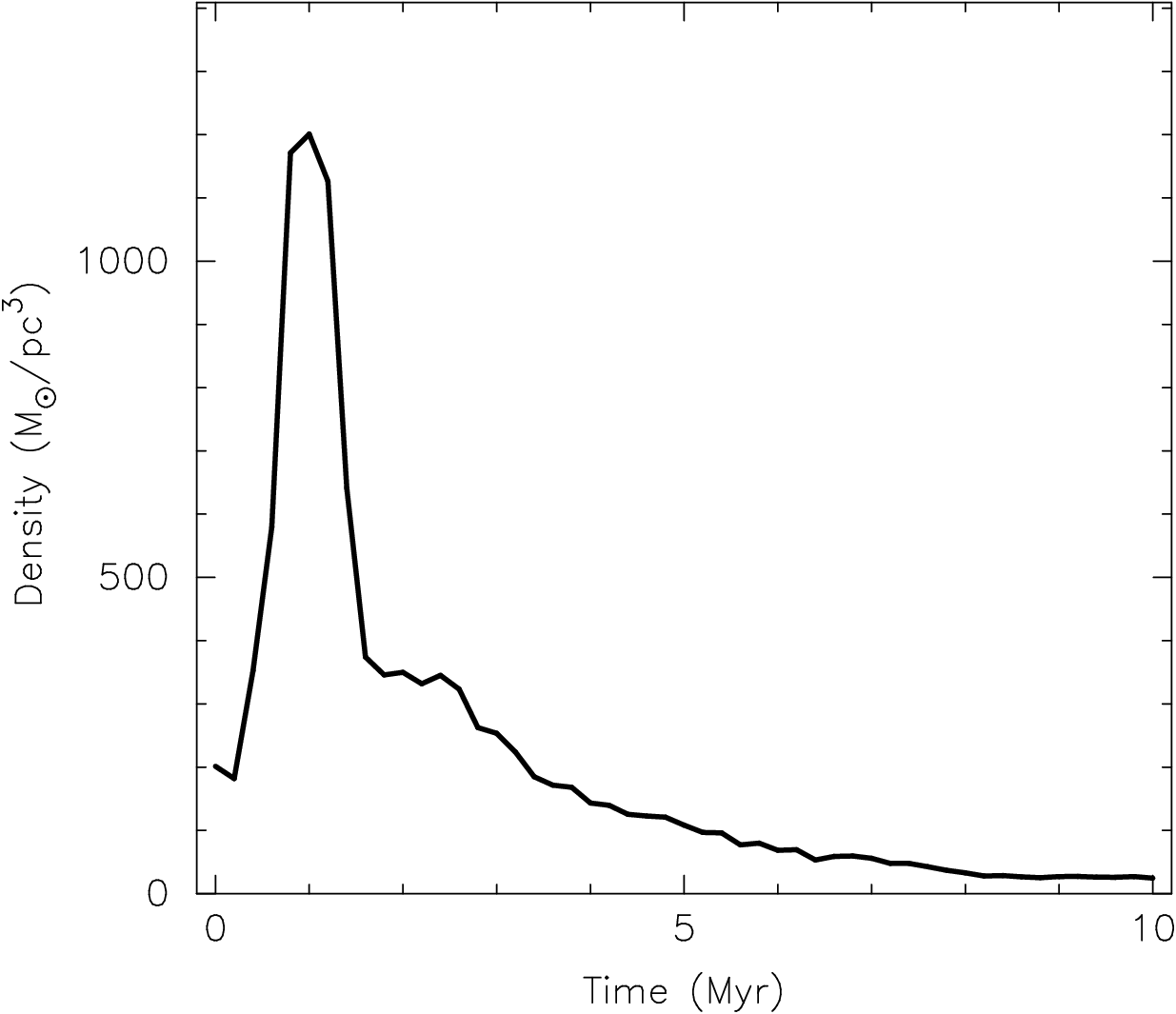}}} 
\caption[bf]{Evolution of (a) the half-mass radius, $r_{1/2}$, and the (b)  central density, $\rho_{\rm cent}$, of the `default' model ($Q = 0.3$, $N_{\rm obj} = 1500$) cluster over the duration of the simulation. The 
cluster reaches its densest phase ($\rho_{\rm cent}$ = 1200\,M$_\odot$\,pc$^{-3}$, $r_{1/2}$ = 0.41\,pc) at 1\,Myr.}
\label{cluster_evolution}
\end{center}
\end{figure*}

We show the evolution over 10\,Myr of the half-mass radius, $r_{1/2}$ of the cluster in Fig.~\ref{evolve-a}, and the central density, $\rho_{\rm cent}$ in Fig.~\ref{evolve-b}. We calculate the half-mass 
radius from the centre of mass of the cluster, and we define the central density as
\begin{equation}
\rho_{\rm cent} = \frac{0.5M_{\rm cluster}}{\frac{4}{3}\pi r_{1/2}^3},
\end{equation}
where $M_{\rm cluster}$ is the total mass of the cluster.

As the cluster collapses, the half-mass radius reaches a minimum of 0.4\,pc, whereas the central density of the cluster rapidly increases in the first Myr, peaking at 1200\,M$_\odot$\,pc$^{-3}$ before the cluster expands and relaxes. The density after 10\,Myr is 24\,M$_\odot$\,pc$^{-3}$. 
Although the virial and supervirial clusters in our analysis do not reach such high densities, we still expect some dynamical interactions within the substructure, as found for clusters with a high stellar binary fraction \citep{Parker11c}.

\begin{figure}
 \begin{center}
\rotatebox{270}{\includegraphics[scale=0.35]{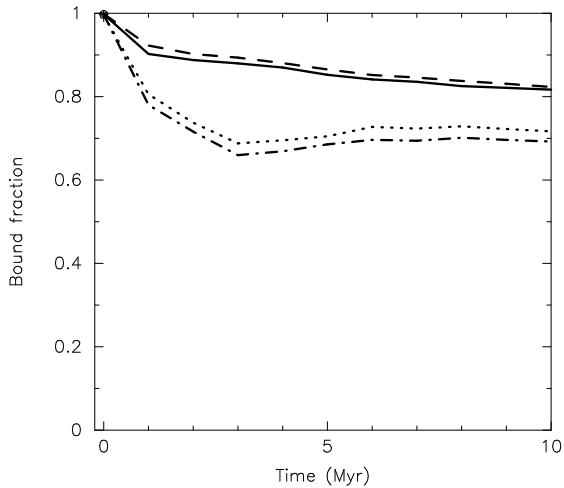}}
\caption[bf]{The fraction of systems bound to the cluster as a function of time. We show the fraction of planets and stars that are energetically bound to the cluster's centre of mass (the solid and dashed 
lines, respectively); and the fraction of planets and stars that are located within two half-mass radii of the cluster centre and therefore would likely be observationally categorised as cluster members (the dot-dashed and 
dotted lines, respectively).}
\label{bound_frac}
\end{center}
\end{figure}

As the cluster evolves, a fraction of planets and stars become unbound from the cluster. We show the fraction of systems that remain bound in Fig.~\ref{bound_frac}. Firstly, we determine whether a planet or star 
is bound to the cluster if it has a negative binding energy with respect to the centre of mass and velocity of the cluster\footnote{If the star or planet is in a binary system then we use the centre of mass velocity of the binary, 
as a planet, or secondary component of a binary may have an orbital velocity in excess of the cluster escape velocity.}. The fraction of such bound planets is shown by the solid line in Fig.~\ref{bound_frac}, and the fraction 
of bound stars is shown by the dashed line. Secondly, we calculate the fractions of planets and stars that reside within two half-mass radii of the centre of the cluster; an observer may not have enough information to determine whether 
an object is energetically bound to the cluster and this second constraint provides a conservative lower limit to the number of `bound' systems. The fractions of planets and stars within $2\,r_{1/2}$ are shown  in Fig.~\ref{bound_frac}  by the dot-dashed, and dotted lines 
respectively. Using this second criteria, we see that at 10\,Myr only 70\,per cent of stars and planets remain within $2\,r_{1/2}$ of the cluster centre. We will refer to systems within $2\,r_{1/2}$ of the cluster centre as `associated' with the cluster, and systems outside of this 
radius as `non-associated'.

\subsection{Liberated and disrupted systems}\label{sec_disrupt}

In our simulations, we determine whether a star or planet is in a bound system using the nearest neighbour method outlined in \citet{Parker09a}. If two stars, or a star 
and a planet, are mutual nearest neighbours, and have a negative binding energy, we deem them to be a bonafide binary system. We also track planets in triple systems; 
if the two stars and the planet, or star and two planets are all mutual nearest and second nearest neighbours, then they are an observed triple. 
Note that such systems are often transient and generally hierarchical, where the inner separation $a_{\rm in}$  is much smaller than the outer separation, $a_{\rm out}$ ($a_{\rm out} >> a_{\rm in}$). 
When this criterion is fulfilled, we determine the binding energy of the inner orbit, and then the outer orbit (using the centre of mass of the inner system). If the outer binding energy is positive 
then we do not include this system as a triple in our analysis.

In the following analysis, we will refer to a planetary system being `disrupted' if it suffers a significant change in either eccentricity or separation (we define what we consider to be reasonable 
thresholds below) but remains bound to its host star, or `liberated' if it is removed from its host star through dynamical interactions. 

Planets that have been liberated can subsequently be captured by a star that is not their parent. Additonally, the interaction of two systems can result in planets being exchanged 
between stars. As these events are both relatively rare, we refer to planets being `captured' if they experience either event. We discuss such systems in Section~\ref{sec_capture}.

\begin{figure*}
 \begin{center}
\setlength{\subfigcapskip}{10pt}
\hspace*{-1.5cm}\subfigure[Planets at 5\,au]{\label{planet_scatter-a}\rotatebox{270}{\includegraphics[scale=0.39]{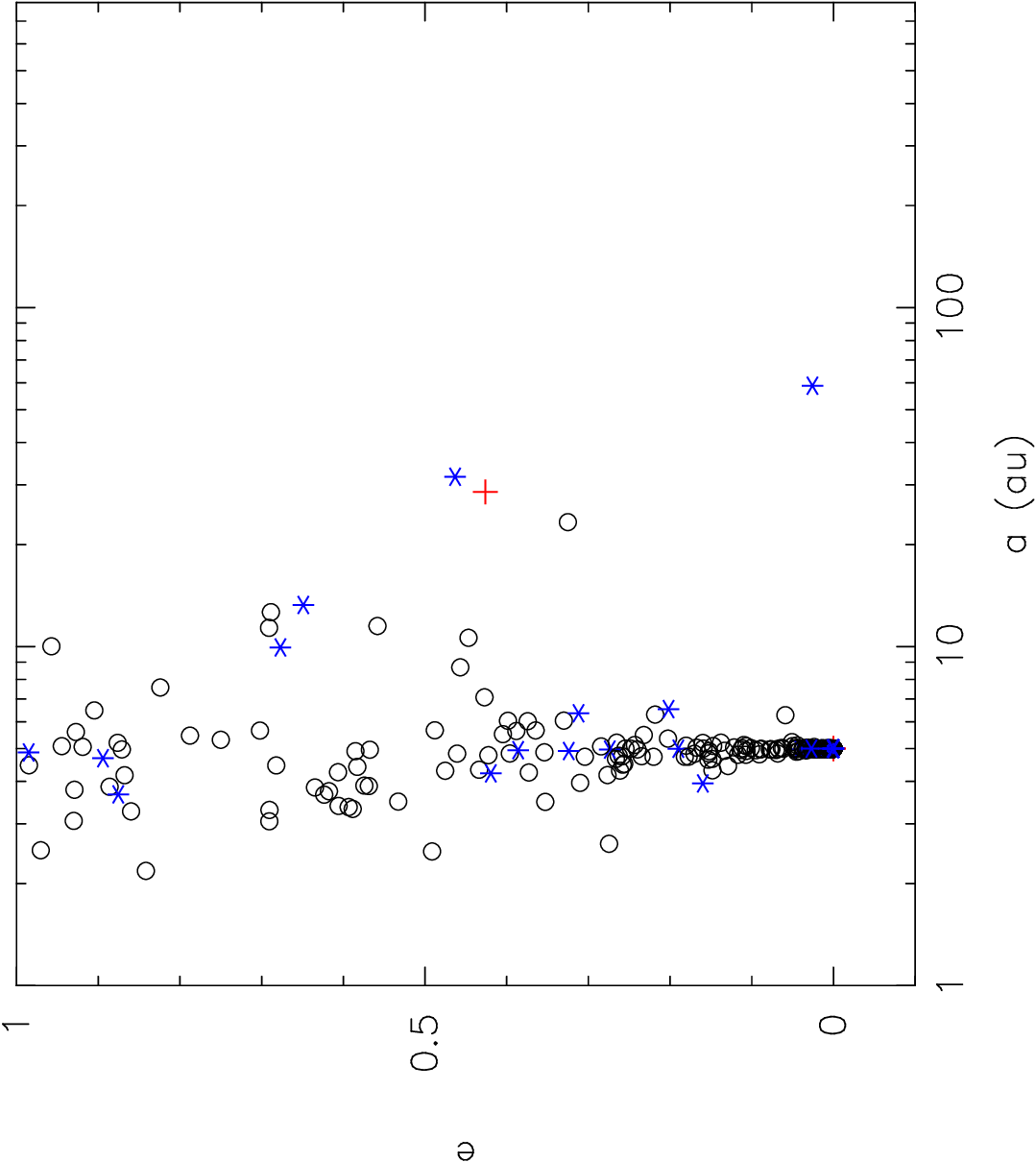}}}
\hspace*{0.3cm} 
\subfigure[Planets at 30\,au]{\label{planet_scatter-b}\rotatebox{270}{\includegraphics[scale=0.39]{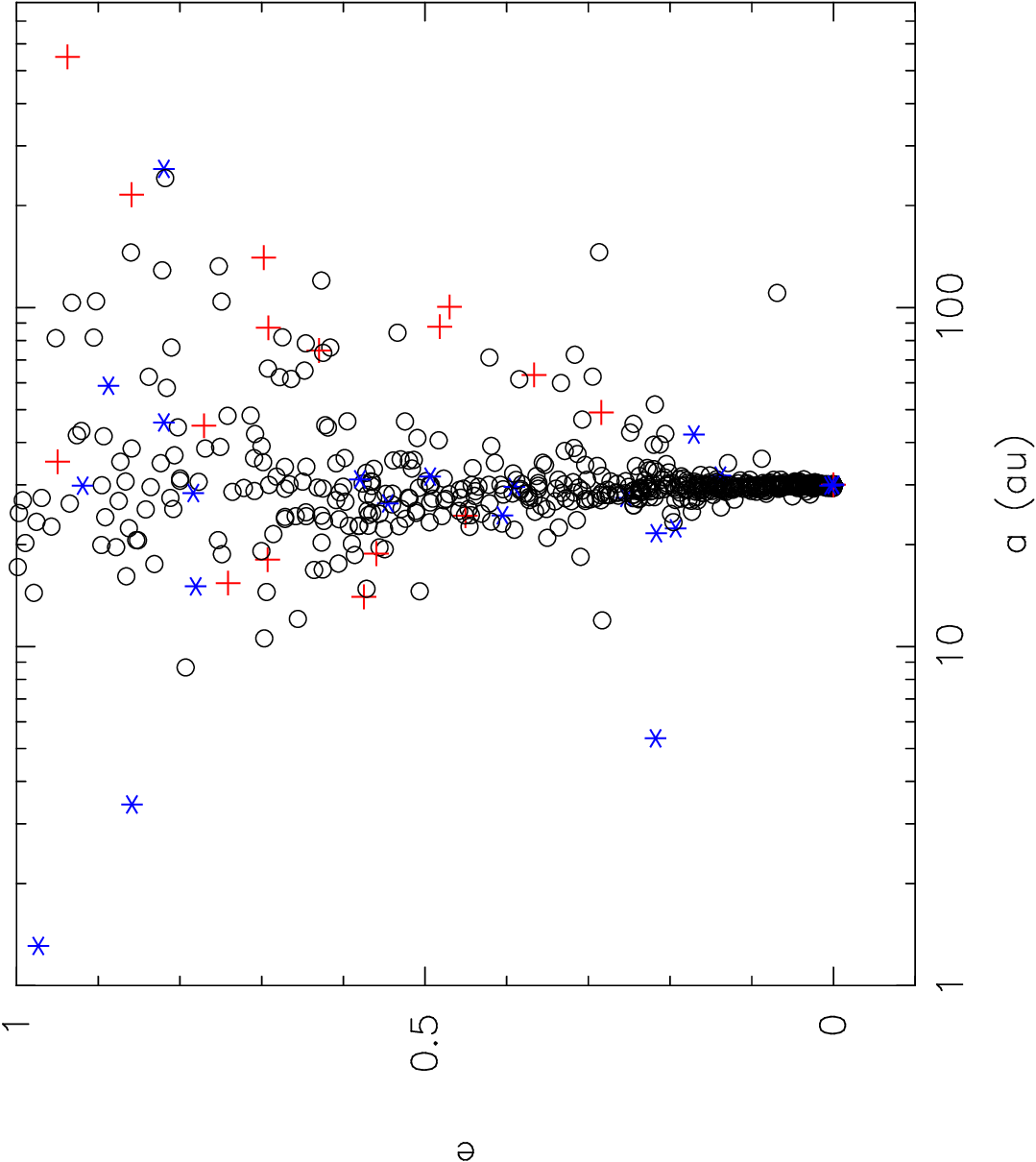}}} 
\caption[bf]{Orbital eccentricity versus separation for 1M$_{\rm Jup}$ planets initially at (a) 5\,au, and (b) 30 \,au, for subvirial $N_{\rm obj} = 1500$ clusters. We have summed together 
the results of 10 different clusters  with the same initial conditions after 10\,Myr of evolution. Circles indicate `birth' planetary systems, crosses show captured planets  
and the asterisks show triple systems containing planets.}
\label{planet_scatter}
\end{center}
\end{figure*}

In Fig.~\ref{planet_scatter} we show the effects of dynamical interactions on planetary orbits in clusters undergoing cool collapse. In this plot we have combined the results of 10 different 
clusters, identical apart from the random number seeds used to initialise the simulations. The reasons for this are two-fold: (i) we improve the statistics (\citet{Adams06} find that averaging 10 simulations 
together produces the required statistical significance), and (ii) the field is the sum of different star formation regions, so once these clusters disperse it becomes difficult to differentiate between 
different formation regions based on the observations of the field alone. In Fig.~\ref{planet_scatter} we plot eccentricity against separation for systems containing planets after 10\,Myr, for 
planets initially at 5\,au (Fig.~\ref{planet_scatter-a}), and planets initially at 30\,au (Fig.~\ref{planet_scatter-b}). The open circles denote primordial planetary systems, 
where the planet is still orbiting its parent star. The crosses indicate captured systems, where the planet is not orbiting its parent star. Finally, the asterisks indicate planets in triple 
systems; the plotted semi-major axis can either be $a_{\rm in}$ or $a_{\rm out}$, depending on whether the triple consists of a star orbiting a star--planet binary, or a planet orbiting a star--star binary, respectively. 
In the rare instance of a triple containing 2 or 3 planets, we plot the inner-most semi-major axis, as this is less susceptible to dynamical break-up.

The orbital eccentricities can be excited from 0 to almost 1 in some cases, and in fewer systems the planetary separations are either hardened or softened. Comparing the systems at 5\,au to those at 30\,au, 
we see that the effect is much more pronounced for the systems at 30\,au.

In order to quantify the effects of cluster evolution on the planets in our simulations, we define two criteria for a planetary system to be `disrupted': either the eccentricity is raised from zero to 0.1, or the semi-major 
axis of the planet decreases or increases by 10 per cent or more. The two processes are by no means mutually exclusive; a system with an altered semi-major axis 
is likely to have a non-zero eccentricity. This eccentricity threshold was chosen in particular because \citet{Raymond11} recently showed that the eccentricity of the innermost giant planet can have significant impact on 
the evolution of forming planetary systems and the chances of survival for terrestrial planets closer to the star. Specifically, these authors modeled the evolution of planetary systems in which the giant planets in the system 
are subject to dynamical instabilities, and they found that in the case of a final eccentricity $\ge 0.1$ for the innermost giant planet, the vast majority of the simulations 
ended up with unstable systems or systems containing only one terrestrial planet. It should be noted that this change in eccentricity would not cause a giant planet at 30\,au to reach the same periapsis as the terrestrial regime of its planetary system; the planet 
would require its eccentricity to be raised to $e \geq 0.85$ for this to occur.

Our threshold for the change in semi-major axis is rather arbitrary; however, as we will see, systems with their eccentricities raised to above 0.1 almost always have their semi-major axis altered by 10\,per cent or more and would therefore 
be `disrupted' based on our first criterion alone.

Adding together the simulations in which we place the planets at 5\,au initially, we find that of the 4123 birth planetary systems, 198 ($5 \pm 2$\,per cent) planets are liberated from their host star, one of which is captured by another star;  
whilst 3925 ($95 \pm 2$\,per cent) survive the 10\,Myr of cluster evolution. Of these 3925 `preserved systems'\footnote{Note that we are not considering systems that formed via capture during the evolution of the cluster.}, 104 ($2.6 \pm 1.3$ per cent) 
have eccentricities excited above 0.1, and 56 ($1.4 \pm 1.0$\,per cent) have their separations altered by more than 10 per cent ($\pm 0.5$\,au).  The number of systems that have their eccentricity \emph{and} semi-major axis altered beyond the thresholds is 55 ($1.4 \pm 1.0$\,per cent).

As is readily apparent from inspection of Fig.~\ref{planet_scatter}, planets on Neptune-like orbits (at 30\,au) are more susceptible to disruption. In this scenario, of the 4148 birth planetary systems,  503 ($12 \pm 3$\,per cent) planets
 are liberated, seventeen of which are captured by another star, whilst 3645 ($88 \pm 3$\,per cent) survive the 10\,Myr of cluster evolution. Of these 3645 preserved systems, 398 ($11 \pm 4$\,per cent) have eccentricities excited above 0.1, and 
189 ($5 \pm 2$\,per cent) have their separations altered by more than 10 per cent ($\pm 3$\,au). 187 ($5 \pm 2$\,per cent) systems have both their eccentricity and semi-major axis altered, suggesting that any system that has its semi-major axis altered 
will also experience a change in eccentricity, whereas a system can suffer a change in eccentricity but its semi-major axis will remain unaffected.   The fraction of planets originally at 30\,au that would directly penetrate the terrestrial regime ($e \geq 0.85$) is $0.8 \pm 0.6$\,per cent.

\begin{figure}
 \begin{center}
\rotatebox{270}{\includegraphics[scale=0.4]{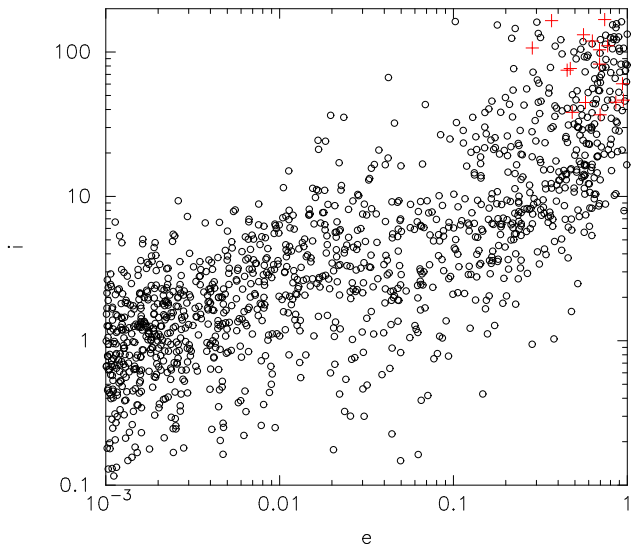}}
\caption[bf]{Inclination in degrees versus eccentricity for all preserved planetary systems (the circles) originally on 30\,au orbits (the planets all have a birth inclination of zero), in our default $N_{\rm obj} = 1500$, $Q = 0.3$ cluster. We also 
show the small number of captured systems by the (red) crosses. The values shown are at time 10\,Myr.}
\label{inc_ecc}
\end{center}
\end{figure}

The inclination angle of a planet's orbit can also be significantly altered during cluster evolution, as shown in Fig.~\ref{inc_ecc}. For the planets originally at 30\,au, we plot inclination against eccentricity, and we see that even planets 
that do not undergo strong changes in eccentricity can have a non-zero inclination\footnote{We also see the same behaviour for the planets originally on 5\,au orbits.}. Most interestingly, planets which experience a significant change in eccentricity (i.e.\,\,$e > 0.1$) 
or are captured tend to have the highest inclinations. We will return to this in Section~\ref{discuss}.

\subsubsection{Dependence on initial virial ratio}

\begin{figure*}
 \begin{center}
\setlength{\subfigcapskip}{10pt}
\hspace*{-0.7cm}\subfigure[Fraction liberated]{\label{vir_rat-a}\rotatebox{270}{\includegraphics[scale=0.27]{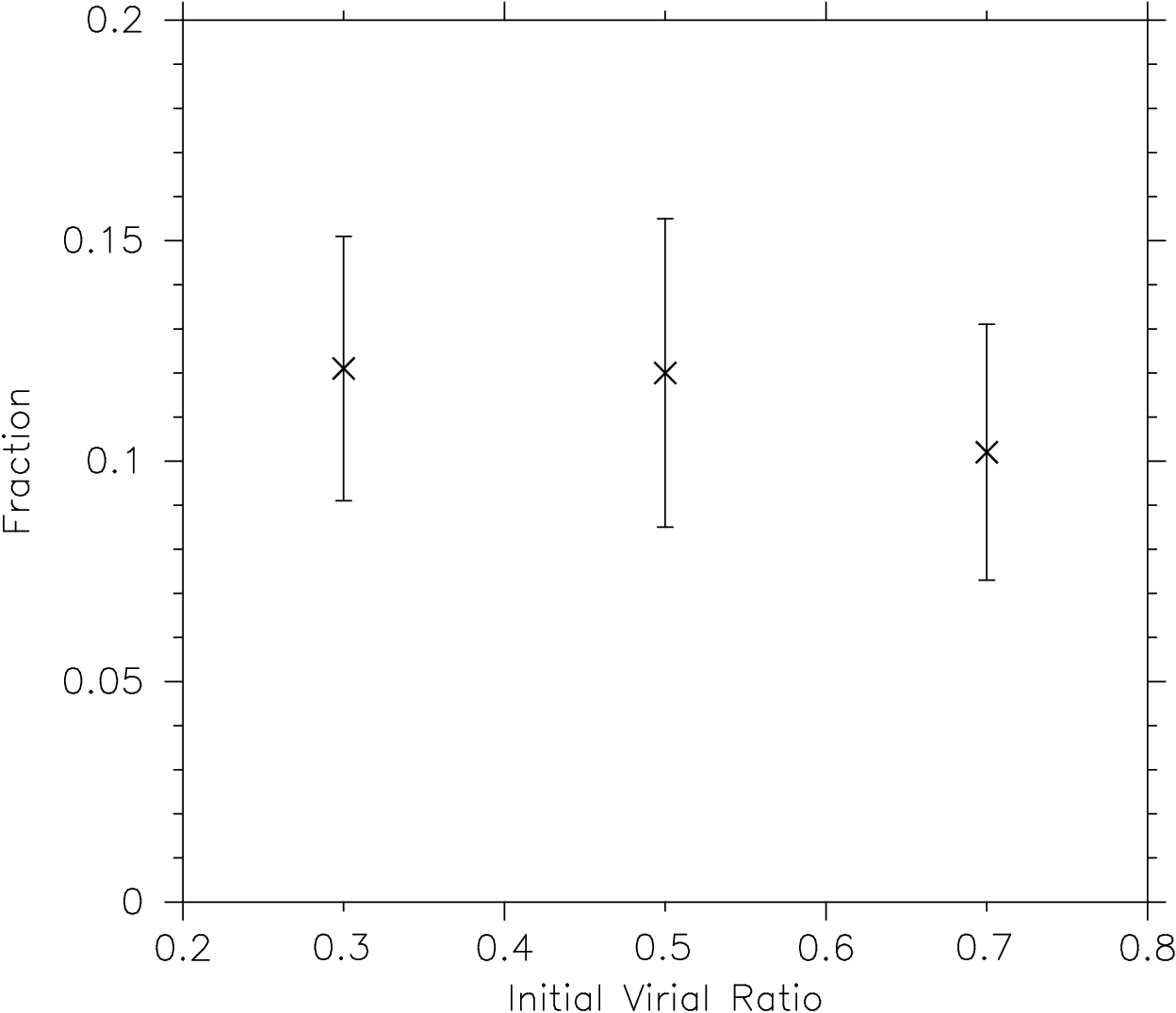}}}
\hspace*{0.3cm} 
\subfigure[Fraction with $e > 0.1$]{\label{vir_rat-b}\rotatebox{270}{\includegraphics[scale=0.27]{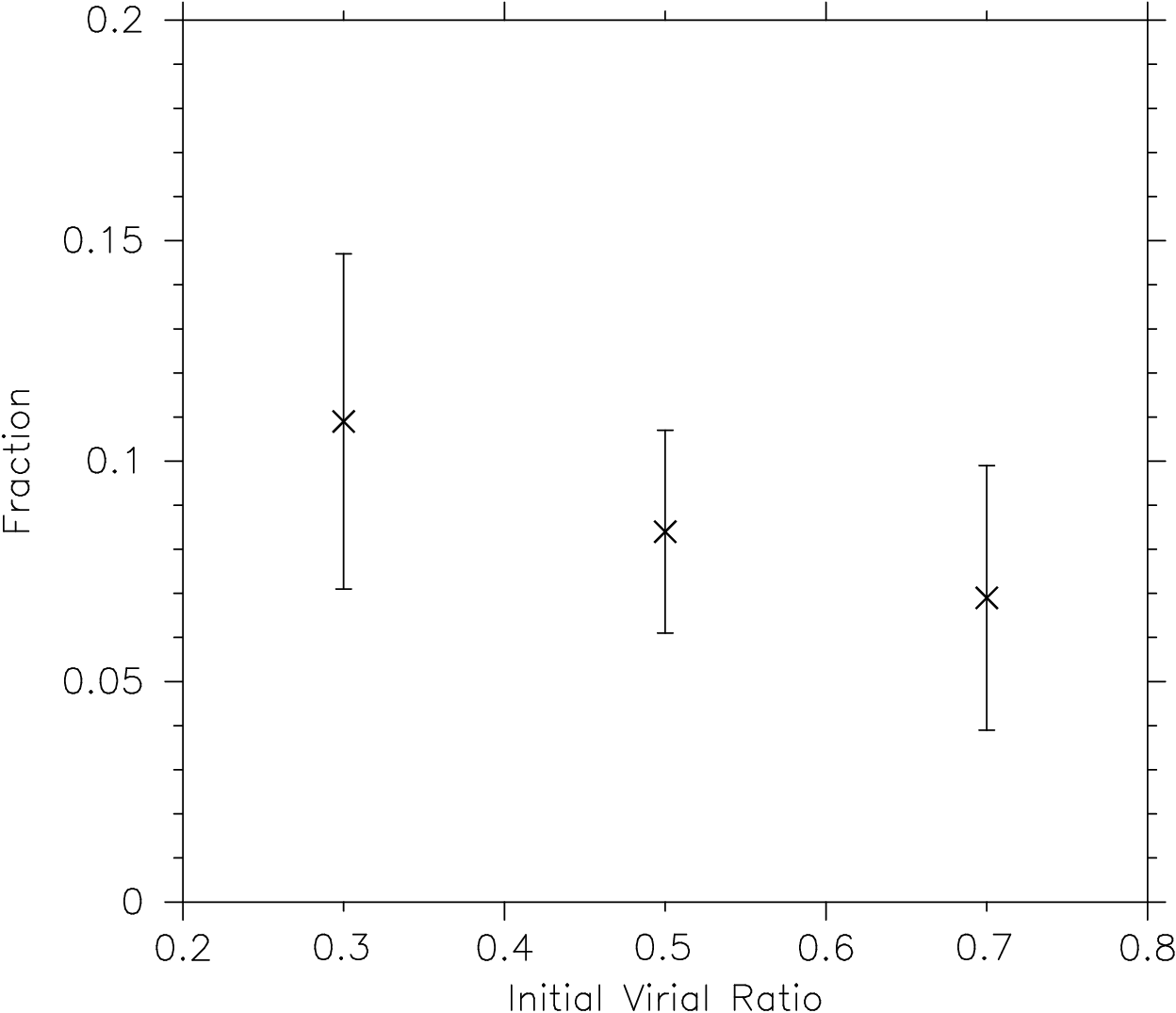}}} 
\hspace*{0.3cm} 
\subfigure[Fraction with $a \pm 10$\,per cent]{\label{vir_rat-c}\rotatebox{270}{\includegraphics[scale=0.27]{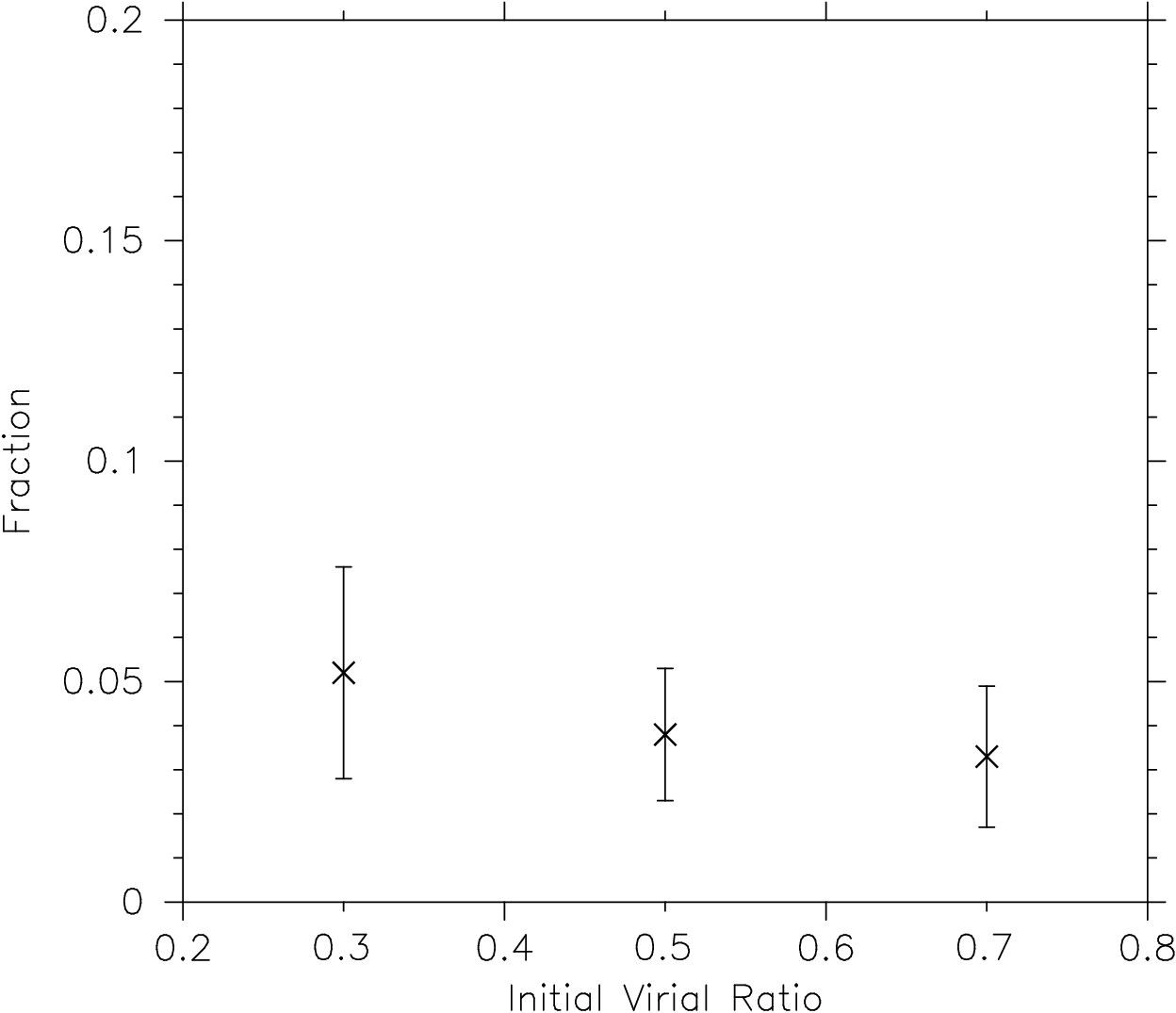}}} 
\caption[bf]{The fraction of (a) liberated systems, (b) preserved systems with eccentricity $ > 0.1$, and (c) preserved systems with semi-major axis altered by more than 10\,per cent, as a function 
of the initial virial ratio of the clusters. We show clusters with initial virial ratio $Q = 0.3$ (subvirial, or `cool', which collapse), $Q = 0.5$ (virial equilibrium) and $Q = 0.7$ (supervirial, or `warm', which expand). 
The clusters have $N_{\rm obj} = 1500$ and the planets were originally on 30\,au orbits. The values shown are at time 10\,Myr.}
\label{virial_ratio}
\end{center}
\end{figure*}

The results described above are for planets in clusters which evolve via the cool collapse scenario. At present, it is unclear what proportion of clusters undergo cool collapse, and the respective proportions 
of clusters that are supervirial and expand from an early age, and those that remain in virial equilibrium. We also conduct simulations in which the overall virial ratio of the cluster was 0.5 (virial equilibrium) and 0.7 (`warm'). 
For simplicity we only consider the clusters with planets initially at 30\,au, and we compare them to the clusters in  cool collapse (virial ratio of 0.3). The results are shown in Fig.~\ref{virial_ratio}. 
In Fig.~\ref{vir_rat-a} we show the fraction of liberated systems at 10\,Myr,  in Fig.~\ref{vir_rat-b} we show 
the fraction of preserved systems that have their eccentricities raised to above 0.1 and in Fig.~\ref{vir_rat-c} we show the fraction of preserved systems that have their separations altered by more than 10 
per cent. 

As we would expect, the clusters undergoing cool collapse (and therefore subject to a very dense phase) are more likely to disrupt planetary systems than clusters that are in virial equilibrium or expanding. One might 
expect that an expanding cluster would not process the planetary population at all; however whilst the cluster is expanding the stars in the subclumps in the fractal interact and decay on a timescale much less 
than the time taken for the cluster to expand. Therefore these systems undergo dynamical interactions, albeit at a lower level than the clusters in virial equilibrium or in cool collapse, as demonstrated in 
Fig.~\ref{virial_ratio}. 


\subsubsection{Dependence on density}

\begin{figure*}
 \begin{center}
\setlength{\subfigcapskip}{10pt}
\hspace*{-0.7cm}\subfigure[Fraction liberated]{\label{dens-a}\rotatebox{270}{\includegraphics[scale=0.27]{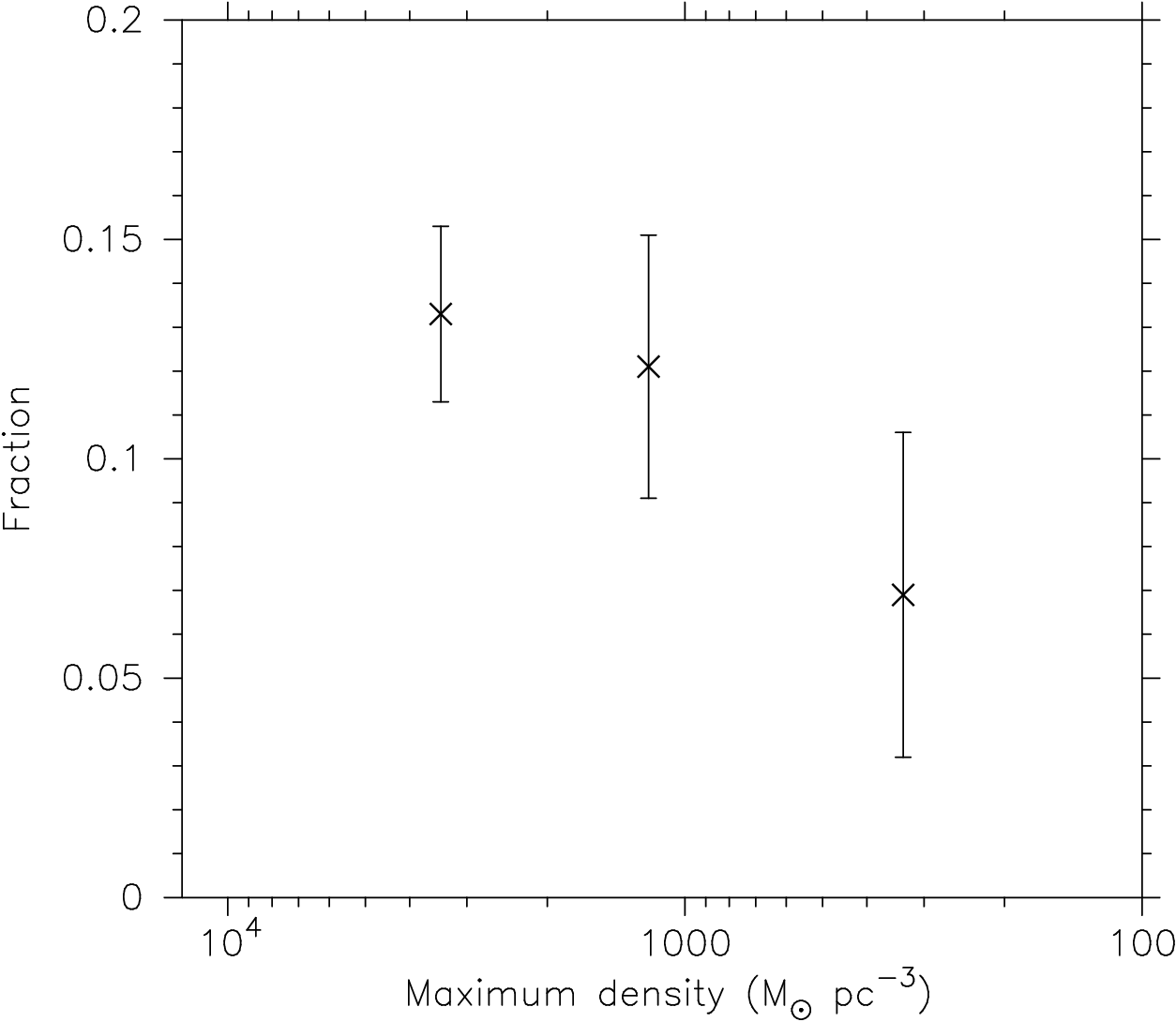}}}
\hspace*{0.3cm} 
\subfigure[Fraction with $e > 0.1$]{\label{dens-b}\rotatebox{270}{\includegraphics[scale=0.27]{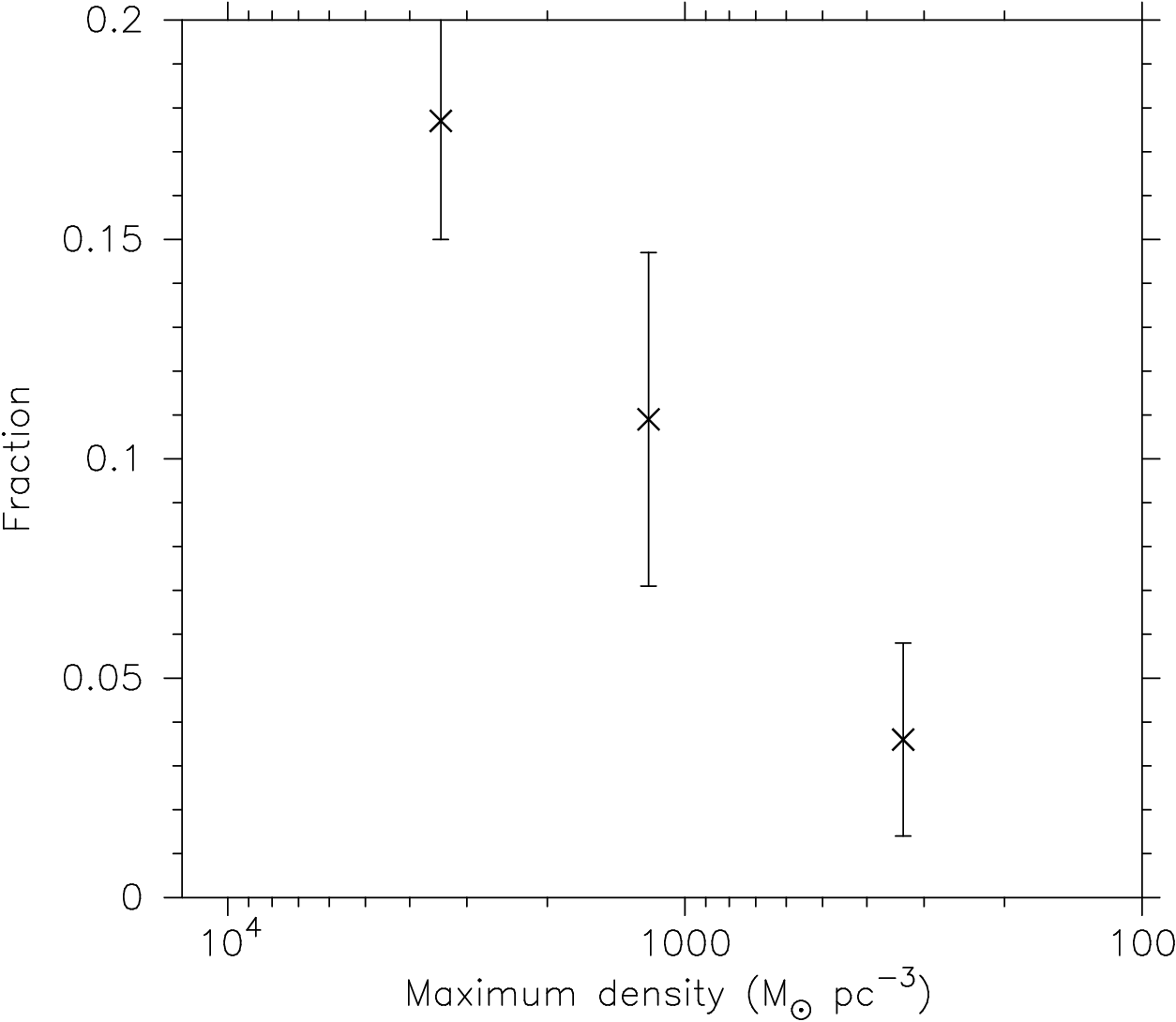}}} 
\hspace*{0.3cm} 
\subfigure[Fraction with $a \pm 10$\,per cent]{\label{dens-c}\rotatebox{270}{\includegraphics[scale=0.27]{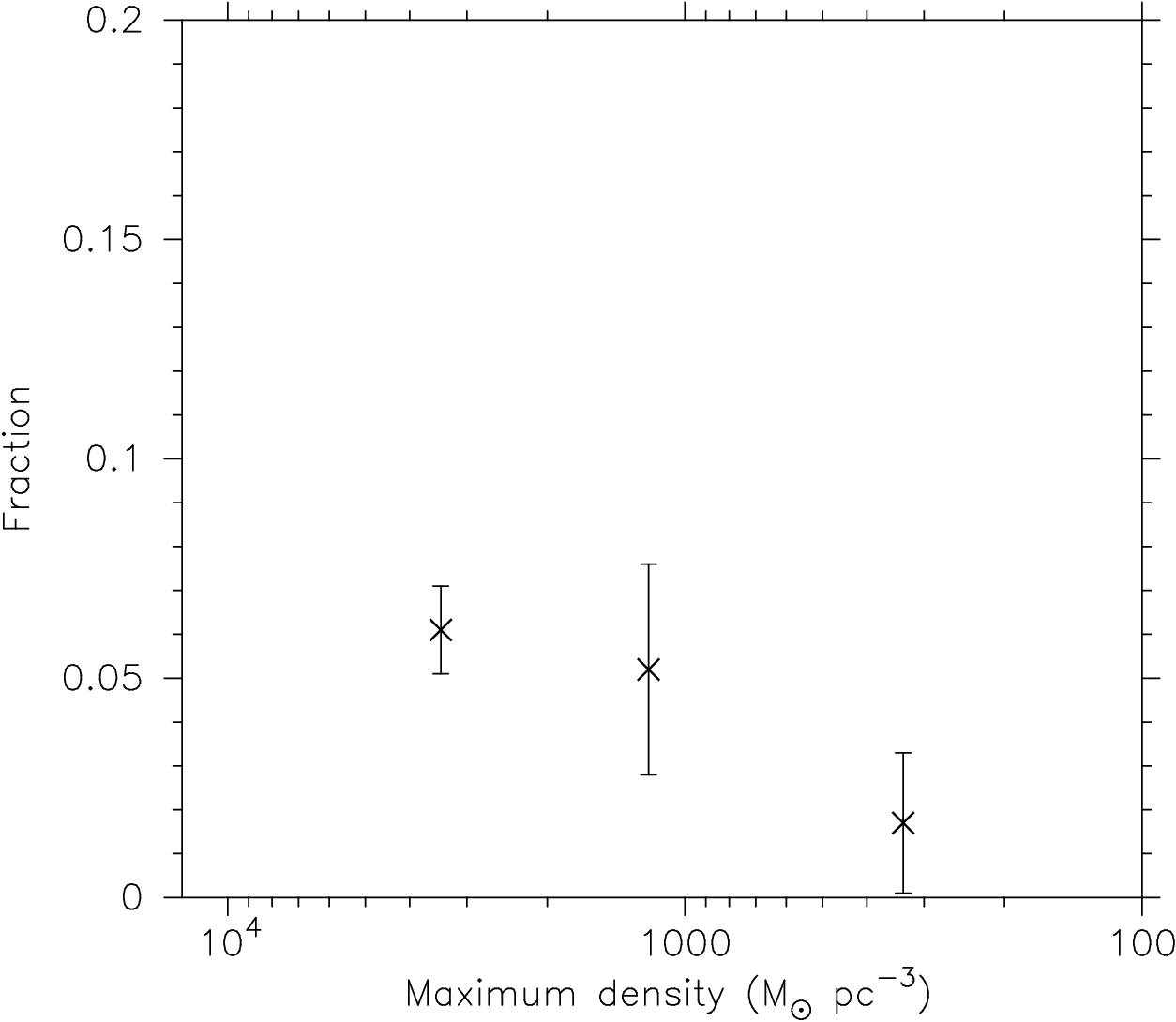}}} 
\caption[bf]{The fraction of (a) liberated systems, (b) preserved systems with eccentricity $ > 0.1$, and (c) preserved systems with semi-major axis altered by more than 10\,per cent, as a function 
of the maximum density reached by the subvirial ($Q = 0.3$) clusters. This corresponds to a time of 0.5\,Myr for clusters containing 3000 objects, 1\,Myr for 1500 objects, and 
2.5\,Myr for the clusters with 200 objects. The values shown are the fractions of liberated/disrupted planets at 10\,Myr.}
\label{density}
\end{center}
\end{figure*}

The fraction of planetary systems that are liberated, and have their eccentricities and semi-major axes altered varies as a function of the density of the cluster. We plot the results (at 10\,Myr) for clusters 
with $Q = 0.3$ and planets originally at 30\,au in Fig.~\ref{density}. In each panel we show the values for clusters with (from left to right) 3000, 1500 and 200 objects. The critical value 
to be considered here is the most dense phase of the cluster's evolution, which occurs at $\sim$ 0.5\,Myr for clusters containing 3000 objects, $\sim$ 1\,Myr for clusters 
containing 1500 objects,  and at $\sim$ 2.5\,Myr for clusters containing only 200 objects. The clusters containing 3000 objects obtain 
maximum densities of $\sim 3420$\,M$_\odot$~pc$^{-3}$ ($\sim$\,7800\,stars\,pc$^{-3}$), whereas at the other end of the scale the clusters containing 200 objects reach densities of 
$\sim 330$\,M$_\odot$~pc$^{-3}$ ($\sim$\,680\,stars\,pc$^{-3}$). In the most dense clusters, the fraction of systems with altered eccentricities is 18 per cent, compared to only 
4 per cent for the least dense clusters. The fraction of systems with altered semi-major axes follows a similar trend, with 6 per cent of systems having altered 
separations in the most dense clusters, compared to only 2 per cent for the least dense clusters.

\subsection{Captured systems}
\label{sec_capture}

As mentioned in Section~\ref{sec_disrupt} there are several systems where planets liberated from their host stars are captured by another star (the crosses in Fig.~\ref{planet_scatter}). 
Whilst there is only one such system for the planets originally at 5\,au, there are seventeen in the 30\,au case. All of these systems are `disrupted' in the sense that their eccentricity is 
$> 0.1$ and their semi-major axis also deviates by more than 10 per cent from the initial value. However, and particularly in the 30\,au case, they do not populate any specific region 
in the plot in Figure~\ref{planet_scatter}, making it observationally impossible to distinguish between captured or just disrupted primordial planets  from the orbital elements alone. 

There is however, a high fraction of captured planets on retrograde orbits. In Fig.~\ref{inc_ecc} we show (by the crosses) the orbital inclinations of planets originally on 30\,au orbits that have been captured by another 
star. Of these seventeen planets, 8 (42\,per cent) are on retrograde orbits ($> 90^\circ$). This compares to the fraction of all planets that are on retrograde orbits of only 1\,per cent.

\subsection{Velocities of liberated planets}

\begin{figure*}
 \begin{center}
\setlength{\subfigcapskip}{10pt}
\hspace*{-1.5cm}\subfigure[Planets liberated from 5\,au orbits within 2\,$r_{1/2}$]{\label{planet_velocities-a}\rotatebox{270}{\includegraphics[scale=0.35]{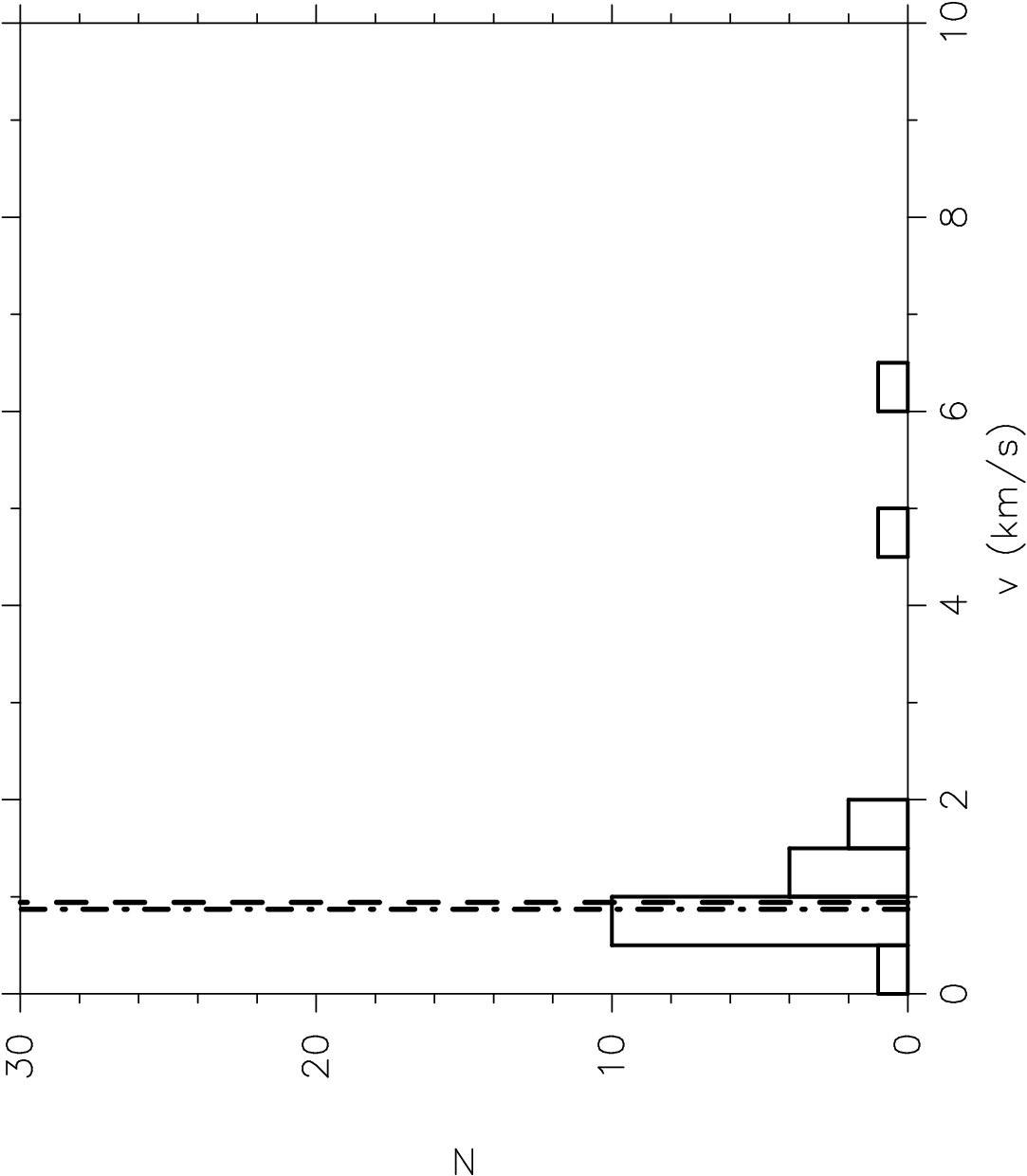}}}
\hspace*{0.3cm} 
\subfigure[Planets liberated from 30\,au orbits within 2\,$r_{1/2}$]{\label{planet_velocities-b}\rotatebox{270}{\includegraphics[scale=0.35]{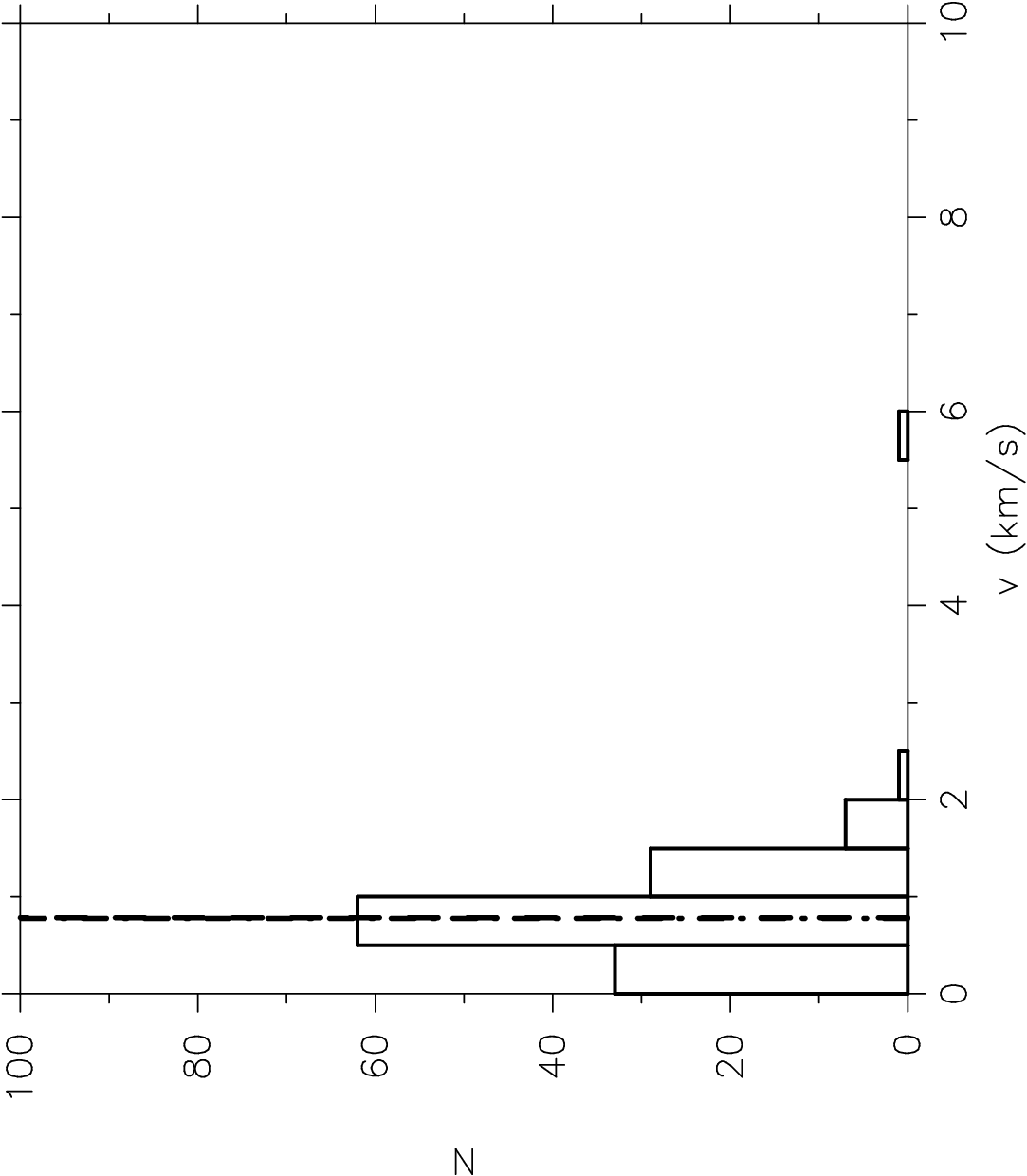}}} 
\hspace*{-1.5cm}\subfigure[Planets liberated from 5\,au orbits outside 2\,$r_{1/2}$]{\label{planet_velocities-c}\rotatebox{270}{\includegraphics[scale=0.35]{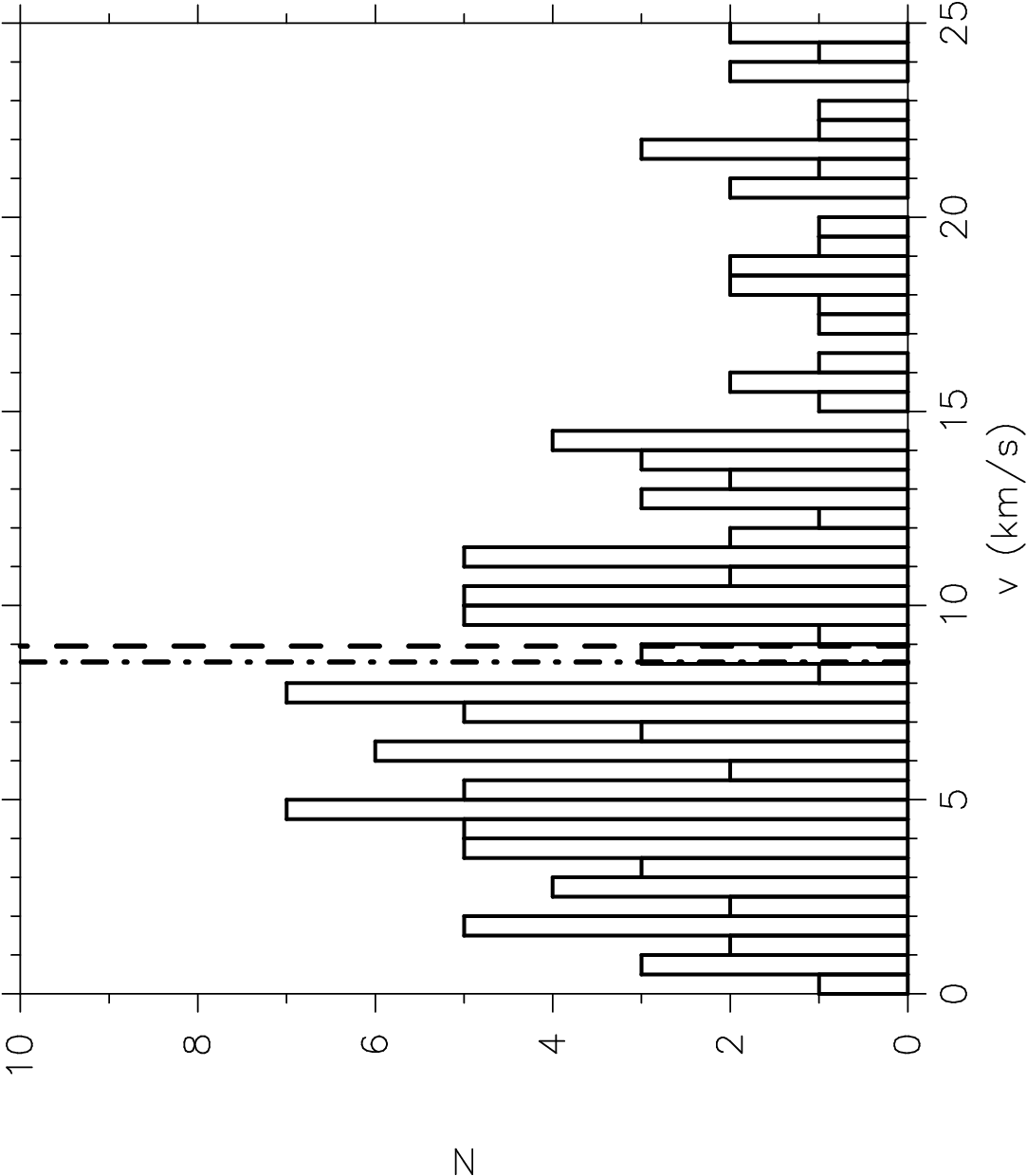}}}
\hspace*{0.3cm} 
\subfigure[Planets liberated from 30\,au orbits outside 2\,$r_{1/2}$]{\label{planet_velocities-d}\rotatebox{270}{\includegraphics[scale=0.35]{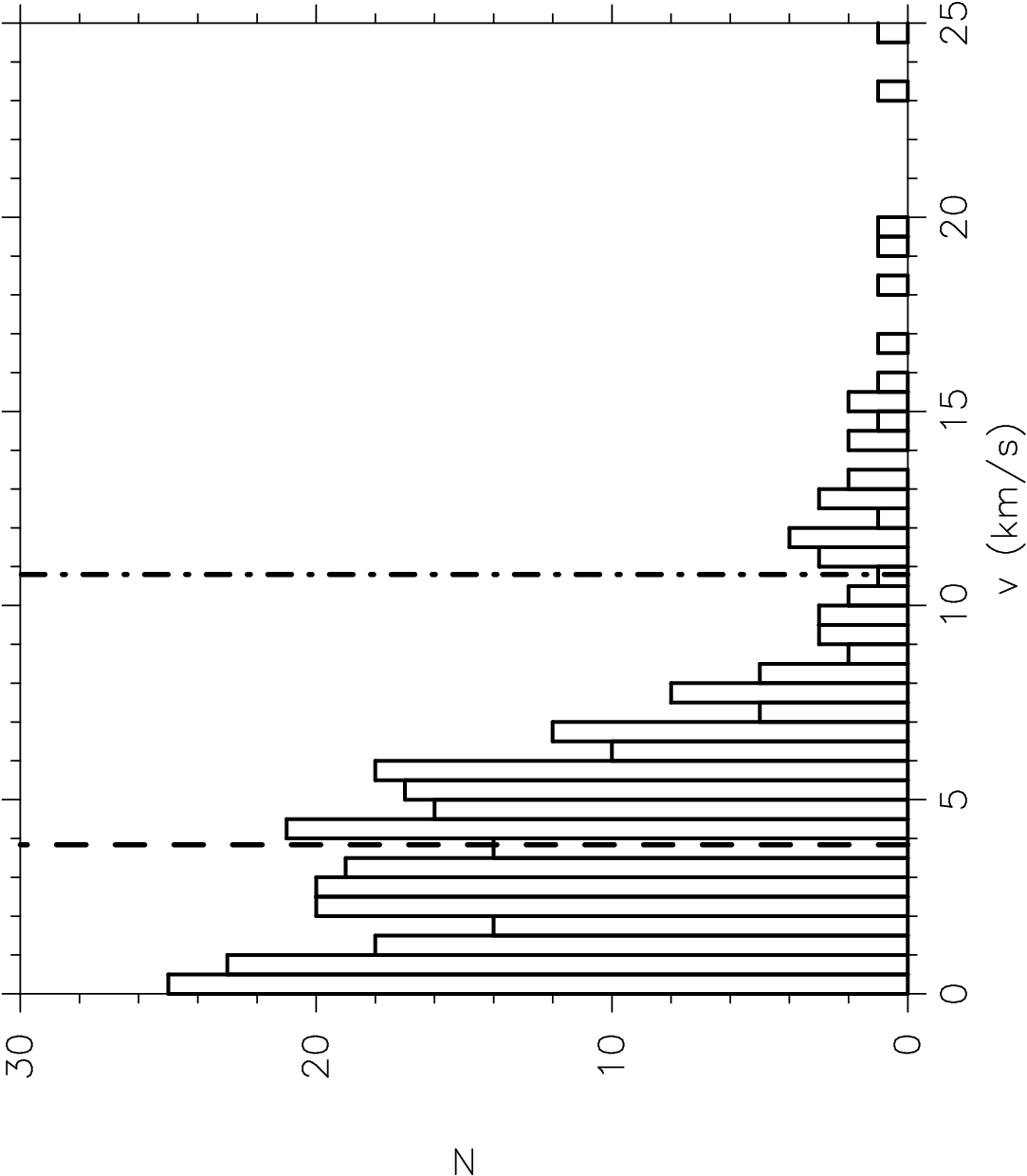}}} 
\caption[bf]{Distribution of velocities of liberated, or free-floating, planets initially at (a) 5\,au, and (b) 30 \,au, which are still associated with the cluster -- i.e.\,\,within two half-mass radii of the centre, for subvirial $N_{\rm obj} = 1500$ clusters. We have summed together 
the results of 10 different clusters  with the same initial conditions after 10\,Myr of evolution. We also show the velocities of planets originally on (c) 5\,au, and (d) 30\,au orbits, that are no longer within two half-mass radii of the cluster centre. The median velocities of 
the liberated planets are shown by the dashed lines. The median velocities of \emph{all} objects (either associated -- panels (a) and (b);  or non-associated -- panels (c) and (d))  are shown by the dot--dashed lines.}
\label{planet_velocities}
\end{center}
\end{figure*}

In Fig.~\ref{planet_velocities} we show the distribution of velocities for planets that are single after 10\,Myr, i.e. those that have been liberated from their host stars 
and are now free-floating. We determine whether a free-floating planet is still associated with the cluster (within two half-mass radii of the cluster 
centre -- see Fig.~\ref{bound_frac}), and then plot the velocity distribution of these single planets in Figs.~\ref{planet_velocities-a}~and~\ref{planet_velocities-b}. 
Fig.~\ref{planet_velocities-a} shows the distribution of planets originally on 5\,au orbits, and Fig.~\ref{planet_velocities-b} shows the distribution of planets originally on 
30\,au orbits. 

We show the median of the free-floating planets' velocity distribution by the dashed line, and the distribution of all of the cluster members by the dot-dashed 
line\footnote{For stars or planets in binary systems, we use the centre-of-mass velocity, as discussed in Section~2.1.}. The planets originally on 5\,au orbits have a median 
velocity of 0.94\,km\,s$^{-1}$, compared to the median cluster velocity of 0.87\,km\,s$^{-1}$. Planets liberated from 30\,au orbits have a median velocity of 0.78\,km\,s$^{-1}$, 
as does the entire cluster.

We also compare the velocity distribution of free-floating planets that are no longer associated with the cluster (at a distance of more than two half-mass radii from the centre) to 
the velocity distribution of every object which is also no longer associated with the cluster. In Fig.~\ref{planet_velocities-c} we show this distribution for planets originally on 5\,au orbits, 
and the corresponding plot for planets originally on 30\,au orbits in Fig.~\ref{planet_velocities-d}. We see that the median non-associated planet velocity (9.0\,km\,s$^{-1}$; the dashed line in the plots) 
is similar to the median non-associated object velocity (8.5\,km\,s$^{-1}$; the dashed line) for planets originally on 5\,au orbits. The median velocity for non-associated planets originally on 30\,au orbits is 
3.8\,km\,s$^{-1}$, whereas the median velocity for all non-associated objects is 10.8\,km\,s$^{-1}$. 

We will discuss these results further in the following Section.



\section{Discussion}
\label{discuss}


From inspection, our results are qualitatively similar to those obtained by \citet{Laughlin98}, who performed cross-sectional scattering experiments on planetary systems assuming an orbital cross section 
for disruption of $\langle\sigma\rangle = (230 {\rm au})^2$, a typical cluster density $n$ of 1000 stars\,pc$^{-3}$ and velocity dispersion $v$ of 1\,kms$^{-1}$. They performed Monte Carlo scattering experiments on 
Jupiter-mass planets at 5\,au with initially zero eccentricity, and noted a spread in orbital parameters similar to our Fig.~\ref{planet_scatter-a}. However, our clusters can reach higher densities (up to $10^4$ stars\,pc$^{-3}$) 
when in cool collapse (Fig.~\ref{density}).  Later authors \citep[e.g.][]{Bonnell01b,Smith01,Adams06} assumed densities consistent with the Orion Nebular cluster (ONC) of $10^3 - 10^4$ stars\,pc$^{-3}$ but with 
smooth and mainly virialised initial conditions. We expand upon this earlier work by looking at the effects of dynamical evolution in substructured, and sub-virial, super-virial and virialised clusters.

The majority of planets remain unscathed during the first 10\,Myr of the cluster's evolution, with typically 90 per cent of planets at 30\,au surviving break-up, and 95 per cent of planets at 5\,au 
surviving. However, of these `preserved' planetary systems, a further 10 per cent would expect to have their eccentricities raised by more than 0.1. The fraction of systems that have their separations 
altered by more than 10 per cent is typically a factor of two lower than the systems that have their eccentricities altered (see Figs.~\ref{virial_ratio}~and~\ref{density}).

Whilst very few planets originally on 5\,au orbits are disrupted, there are a number that have high eccentricities. Quite possibly, these planets will eventually be ejected from the system, as they are likely to 
cross the orbits of other planets in the system, leading to planet--planet scattering events \citep*[e.g.][and references therein]{Malmberg07a}. It is unlikely that inner terrestrial planets would be preserved 
or even able to form if a Jupiter-mass planet at 5\,au had a large orbital eccentricity \citep{Raymond11}.

Of the larger fraction of planets originally at 30\,au that have their separations and eccentricities increased, there is a significant population that occupy a phase space in which they might 
be directly detected in future imaging surveys carried out with dedicated planet-finding instruments such as Spectro-Polarimetric High-contrast Exoplanet REsearch (SPHERE) at the VLT \citep{Beuzit06} and Gemini Planet Imager (GPI) at the Gemini Observatory 
\citep{Macintosh06}. In particular the planets with high eccentricities and large semi-major axes spend a significant proportion of their orbital period far away from their host star. Surveys carried out 
in recent years are yet to reach the required sensitivity to detect objects with masses below a few Jupiter masses \citep[e.g.,][]{Lafreniere07,Chauvin10}, but this may change with the next 
generation of instruments. If a population of distant planets is detected and if formation mechanisms preclude the formation in-situ of planets at such separations, then their origin 
could be dynamical. We note that a  significant number of systems (25 per cent) with $e > 0.3$ and $a > 50$\,au are actually captured systems, implying that planets observed with these orbital 
characteristics may not be orbiting their parent star. This possibility should also be kept in mind when the origin of the distant exoplanets Fomalhaut b \citep{Kalas08} and 1RXS J1609 b 
\citep{Lafreniere10} with projected separations of $\sim$115 au and $\sim$330 au, respectively, is discussed.

A large number of preserved systems have inclination angles raised due to interactions in the cluster (see Fig.~\ref{inc_ecc}). Whilst the change in angle (from $i = 0^\circ$ at birth) for the majority of these systems 
is only of order a few degrees, systems which we classify as being dynamically disrupted (based on $e > 0.1$) are likely to have inclination angles in the range where they could be lead to secondary dynamical processes, 
such as the  Kozai mechanism \citep{Kozai62}. The Kozai mechanism can operate if the inclination angle of an outer body is raised to within the range $39.23^\circ < i_{\rm Koz} < 140.77^\circ$ of the orbital plane 
of an inner body. The total fraction of preserved systems that have their 
angles excited to this range is only 2\,per cent; however, for the systems with $e > 0.1$ the fraction is much higher, at 23\,per cent. Additionally, of the captured systems, 76\,per cent have inclination angles in the Kozai range.

Recently, \citet{Naoz11} showed that Hot Jupiters, i.e., gas giant planets in a very close-in prograde or retrograde orbit, may result from secular planet-planet interactions. 
In their simulations a gas giant planet is initially orbiting on an almost circular orbit at several au, while a second gas giant planet is orbiting on a distant (a few tens of au), eccentric, and highly inclined orbit. 
Angular momentum exchange and eventually tidal dissipation may then lead to eccentricity fluctuations, orbital decay and circularisation of the first planet. In some cases even the orbital inclination was 
significantly altered leading to a retrograde orbital motion. Our results show that dynamical interactions taking place in the cluster phase are one possibility to provide the initial conditions for this mechanism to work\footnote{An alternative mechanism to provide these initial conditions, 
where gas is captured onto a protoplanetary disc and alters the inclination angle of the disc, was recently proposed by \citet{Thies11}.}. 
Therefore, in a more general sense, the altered eccentricities, semi-major axes and inclinations observed in our simulations may not be representative of the dynamical end-state of these planetary systems. 

The fraction of planetary systems that are affected by direct dynamical processing is dependent on density but also on the initial virial ratio, $Q$, of the system. \citet{Allison09b} and \citet{Parker11c} have 
shown that the `cool collapse' scenario can explain the observed level of mass segregation, and the binary separation distribution in the ONC, if the cluster 
is initially substructured. 
If a significant proportion of clusters evolve in this fashion, then we would expect a significant fraction (12\,per cent) of free-floating planets per cluster, and 10 per cent of the planets to have altered orbital parameters. This compares to 
around 10 per cent of extrasolar planets that could be affected by secondary dynamical effects (such as the Kozai mechnism) in a dense cluster environment \citep{Parker09c}. If the cluster is initially in virial equilibrium, or is supervirial, fewer planetary systems are 
affected, although the clumps in the substructure dynamically decay in all cases, meaning that planets will be affected in almost all clustered environments.


The respective velocity distributions of the free-floating planets warrant special mention. Planets that are free-floating, but still observationally associated with the cluster, have a very similar median velocity to the whole cluster. One may 
naively conclude from this that observing a similar velocity distribution for free-floating planetary mass objects compared to stars would indicate that they represent the very low-mass tail of the IMF, when in actual fact they have been 
liberated from a planetary system. Conversely, planets liberated from 5\,au orbits which are not associated with the cluster appear to have very similar velocities to stars, whereas planets originally at 30\,au have lower velocities. So, 
in principle one could observe different velocities for planets on the periphery of star forming regions and erroneously conclude that their formation scenarios were different. Several studies have already identified potential candidates for free-floating planetary mass 
objects in young star-forming regions, e.g., in $\sigma$ Orionis \citep{Bihain09}, the Orion Nebular Cluster \citep{Weights09}, and Ophiuchus \citep{Harvey10}. Thus far no information about the velocities of these objects is available but once this is the case our 
results suggest that the analysis of free-floating planets' velocities should not be used to draw direct conclusions about their formation scenarios.


\section{Conclusions}
\label{conclude}

Observations suggest that the initial conditions of star clusters are cool and substructured. Few authors have examined the effects of dynamical evolution in clusters on planets through direct $N$-body simulations,  
and none have considered initially substructured environments. We have conducted $N$-body simulations of star clusters in which we place a single Jupiter-mass planet at 5\,au 
or 30\,au around roughly half of the stars in the cluster. The remainder of stars have a stellar binary companion, as a high binary fraction is observed in most star-forming regions. We have kept the level of substructure 
constant, adopting a fractal distribution with fractal dimension $D = 2.0$. This gives a moderate amount of substructure, although we note that some star formation regions may have even more primordial 
substructure \citep{Cartwright04,Sanchez09,Schmeja11}. 

We have varied the number of objects in the cluster, adopting $N_{\rm obj} = 200, N_{\rm obj} = 1500$ or $N_{\rm obj} = 3000$.  We also vary the initial virial ratio, $Q$, creating `cool' ($Q = 0.3$), virialised  
($Q = 0.5$) and `warm' ($Q = 0.7$) clusters. We dynamically evolve each cluster for 10\,Myr and examine the effects of dynamical evolution on the planetary systems. Our conclusions can be summarised as 
follows:

(i) In our reference case ($Q = 0.3$), we find that  $\sim$10 per cent of planets at 30\,au are liberated from their host stars via interactions in the cluster. Of the planets that are `preserved', another $\sim$10 per cent have their eccentricities altered by more 
than 0.1 from an initially circular orbit ($e = 0$). A smaller fraction (typically 5 per cent) have their orbital separations altered by more than 3\,au. The respective numbers for planets originally at 5\,au is lower, 
typically by a factor of two. 

(ii) The planets with increased separation \emph{and} eccentricity would be candidates for future direct imaging searches as they spend a large fraction of their orbit at large distances from their host star. However, 
many planets (25\,per cent) in the $e > 0.3, a > 50$\,au phase space are not primordial systems, and have formed via exchange interactions or capture in the cluster.

(iii) The fractions of `liberated' and `disrupted' planets depend on the initial virial ratio of the cluster. If the cluster is in cool collapse, it reaches a much denser state than a cluster that is in virial equilibrium or 
expanding. However, planets in all clusters are affected by disruption due to low-N dynamical decay and regions of localised high density in the substructure.

(iv) A significant fraction of the `disrupted' planets also suffer a significant change in the orbital inclination which may lead to further dynamical perturbations with other planetary bodies in these systems such as 
secular planet-planet interactions or even planet-planet scattering. The former process may then even lead to the creation of close-in hot Jupiters with retrograde orbits.


(v) Planets that are captured during the evolution of the cluster are indistinguishable from disrupted planets in terms of their eccentricity and semi-major axis, but such planets are more likely to be on retrograde orbits, with an inclination angle $> 90^\circ$.

(vi) The distribution of velocities of liberated planets which remain associated with the cluster are similar to the stellar velocities, irrespective of the planet's original semi-major axis. This suggests that 
planets with similar velocities to stars could be mistaken for a member of the low-mass tail of the IMF which formed from core collapse, when in fact their formation history was very different. Planets that are not associated with 
 the cluster tend to have similar velocities to the median non-associated stellar velocity if they formed at 5\,au, whereas planets originally on 30\,au orbits have a median velocity that is much lower than the 
non-associated stellar velocity. 

In a future paper we will expand this work to study the effects of dynamical evolution in clusters in which planets have formed in primordial stellar binary systems.

\section*{Acknowledgments}

We thank the anonymous referee for their insightful comments and suggestions on the original manuscript, which have lead to a significantly improved paper. We also thank Simon Goodwin for helpful discussions.
The simulations in this work were performed on the \texttt{BRUTUS} computing cluster at ETH Z{\"u}rich.

\bibliographystyle{mn2e}
\bibliography{ref_cluster_planets}

\begin{thebibliography}{}

\bibitem[\protect\citeauthoryear{Abt, Gomez \& Levy}{Abt et~al.}{1990}]{Abt90}
Abt H.~A.,  Gomez A.~E.,    Levy S.~G.,  1990, ApJS, 74, 551

\bibitem[\protect\citeauthoryear{Adams, Proszkow, Fatuzzo \& Myers}{Adams
  et~al.}{2006}]{Adams06}
Adams F.~C.,  Proszkow E.~M.,  Fatuzzo M.,    Myers P.~C.,  2006, ApJ, 641, 504

\bibitem[\protect\citeauthoryear{Allison \& Goodwin}{Allison \&
  Goodwin}{2011}]{Allison11}
Allison R.~J.,  Goodwin S.~P.,  2011, MNRAS, 415, 1967

\bibitem[\protect\citeauthoryear{Allison, Goodwin, Parker, de Grijs, {Portegies
  Zwart} \& Kouwenhoven}{Allison et~al.}{2009}]{Allison09b}
Allison R.~J.,  Goodwin S.~P.,  Parker R.~J.,  de Grijs R.,  {Portegies Zwart}
  S.~F.,    Kouwenhoven M. B.~N.,  2009, ApJ, 700, L99

\bibitem[\protect\citeauthoryear{Allison, Goodwin, Parker, {Portegies Zwart} \&
  de Grijs}{Allison et~al.}{2010}]{Allison10}
Allison R.~J.,  Goodwin S.~P.,  Parker R.~J.,  {Portegies Zwart} S.~F.,    de
  Grijs R.,  2010, MNRAS, 407, 1098

\bibitem[\protect\citeauthoryear{{Beuzit}, {Feldt}, {Dohlen}, {Mouillet},
  {Puget}, {Antichi}, {Baruffolo}, {Baudoz}, {Berton}, {Boccaletti},
  {Carbillet}, {Charton}, {Claudi}, {Downing} \& {Feautrier}}{{Beuzit}
  et~al.}{2006}]{Beuzit06}
{Beuzit} J.-L.,  {Feldt} M.,  {Dohlen} K.,  {Mouillet} D.,  {Puget} P.,
  {Antichi} J.,  {Baruffolo} A.,  {Baudoz} P.,  {Berton} A.,  {Boccaletti} A.,
  {Carbillet} M.,  {Charton} J.,  {Claudi} R.,  {Downing} M.,    {Feautrier} P.
  e.~a.,  2006, The Messenger, 125, 29

\bibitem[\protect\citeauthoryear{{Bihain}, {Rebolo}, {Zapatero Osorio},
  {B{\'e}jar}, {Vill{\'o}-P{\'e}rez}, {D{\'{\i}}az-S{\'a}nchez},
  {P{\'e}rez-Garrido}, {Caballero}, {Bailer-Jones} \& {Barrado y
  Navascu{\'e}s}}{{Bihain} et~al.}{2009}]{Bihain09}
{Bihain} G.,  {Rebolo} R.,  {Zapatero Osorio} M.~R.,  {B{\'e}jar} V.~J.~S.,
  {Vill{\'o}-P{\'e}rez} I.,  {D{\'{\i}}az-S{\'a}nchez} A.,  {P{\'e}rez-Garrido}
  A.,  {Caballero} J.~A.,  {Bailer-Jones} C.~A.~L.,    {Barrado y
  Navascu{\'e}s} D. e.~a.,  2009, A\&A, 506, 1169

\bibitem[\protect\citeauthoryear{Bonnell, Smith, Davies \& Horne}{Bonnell
  et~al.}{2001}]{Bonnell01b}
Bonnell I.~A.,  Smith K.~W.,  Davies M.~B.,    Horne K.,  2001, MNRAS, 322, 859

\bibitem[\protect\citeauthoryear{Bressert, Bastian, Gutermuth, Megeath, Allen,
  {Evans, II}, Rebull, Hatchell, Johnstone, Bourke, Cieza, Harvey, Merin, Ray
  \& Tothill}{Bressert et~al.}{2010}]{Bressert10}
Bressert E.,  Bastian N.,  Gutermuth R.,  Megeath S.~T.,  Allen L.,  {Evans,
  II} N.~J.,  Rebull L.~M.,  Hatchell J.,  Johnstone D.,  Bourke T.~L.,  Cieza
  L.~A.,  Harvey P.~M.,  Merin B.,  Ray T.~P.,    Tothill N. F.~H.,  2010,
  MNRAS, 409, L54

\bibitem[\protect\citeauthoryear{Cartwright \& Whitworth}{Cartwright \&
  Whitworth}{2004}]{Cartwright04}
Cartwright A.,  Whitworth A.~P.,  2004, MNRAS, 348, 589

\bibitem[\protect\citeauthoryear{{Chauvin}, {Lagrange}, {Bonavita},
  {Zuckerman}, {Dumas}, {Bessell}, {Beuzit}, {Bonnefoy}, {Desidera}, {Farihi},
  {Lowrance}, {Mouillet} \& {Song}}{{Chauvin} et~al.}{2010}]{Chauvin10}
{Chauvin} G.,  {Lagrange} A.-M.,  {Bonavita} M.,  {Zuckerman} B.,  {Dumas} C.,
  {Bessell} M.~S.,  {Beuzit} J.-L.,  {Bonnefoy} M.,  {Desidera} S.,  {Farihi}
  J.,  {Lowrance} P.,  {Mouillet} D.,    {Song} I.,  2010, A\&A, 509, A52+

\bibitem[\protect\citeauthoryear{Crowther, Schnurr, Hirschi, Yusof, Parker,
  Goodwin \& Kassim}{Crowther et~al.}{2010}]{Crowther10}
Crowther P.~A.,  Schnurr O.,  Hirschi R.,  Yusof N.,  Parker R.~J.,  Goodwin
  S.~P.,    Kassim H.~A.,  2010, MNRAS, 408, 731

\bibitem[\protect\citeauthoryear{Davies \& Sigurdsson}{Davies \&
  Sigurdsson}{2001}]{Davies01}
Davies M.~B.,  Sigurdsson S.,  2001, MNRAS, 324, 612

\bibitem[\protect\citeauthoryear{Duquennoy \& Mayor}{Duquennoy \&
  Mayor}{1991}]{Duquennoy91}
Duquennoy A.,  Mayor M.,  1991, A\&A, 248, 485

\bibitem[\protect\citeauthoryear{Fabrycky \& Tremaine}{Fabrycky \&
  Tremaine}{2007}]{Fabrycky07}
Fabrycky D.,  Tremaine S.,  2007, ApJ, 669, 1298

\bibitem[\protect\citeauthoryear{Figer}{Figer}{2005}]{Figer05}
Figer D.~F.,  2005, Nature, 434, 192

\bibitem[\protect\citeauthoryear{Fischer \& Marcy}{Fischer \&
  Marcy}{1992}]{Fischer92}
Fischer D.~A.,  Marcy G.~W.,  1992, ApJ, 396, 178

\bibitem[\protect\citeauthoryear{Fregeau, Chatterjee \& Rasio}{Fregeau
  et~al.}{2006}]{Fregeau06}
Fregeau J.~M.,  Chatterjee S.,    Rasio F.~A.,  2006, ApJ, 640, 1086

\bibitem[\protect\citeauthoryear{Goodwin}{Goodwin}{2010}]{Goodwin10}
Goodwin S.~P.,  2010, {Royal Society of London Philosophical Transactions
  Series A}, 368, 851

\bibitem[\protect\citeauthoryear{Goodwin \& Whitworth}{Goodwin \&
  Whitworth}{2004}]{Goodwin04a}
Goodwin S.~P.,  Whitworth A.~P.,  2004, A\&A, 413, 929

\bibitem[\protect\citeauthoryear{{Harvey}, {Jaffe}, {Allers} \& {Liu}}{{Harvey}
  et~al.}{2010}]{Harvey10}
{Harvey} P.~M.,  {Jaffe} D.~T.,  {Allers} K.,    {Liu} M.,  2010, ApJ, 720,
  1374

\bibitem[\protect\citeauthoryear{Heggie}{Heggie}{1975}]{Heggie75}
Heggie D.~C.,  1975, MNRAS, 173, 729

\bibitem[\protect\citeauthoryear{Holman, Touma \& Tremaine}{Holman
  et~al.}{1997}]{Holman97}
Holman M.,  Touma J.,    Tremaine S.,  1997, Nature, 386, 254

\bibitem[\protect\citeauthoryear{Hurley \& Shara}{Hurley \&
  Shara}{2002}]{Hurley02}
Hurley J.~R.,  Shara M.~M.,  2002, ApJ, 565, 1251

\bibitem[\protect\citeauthoryear{Innanen, Zheng, Mikkola \& Valtonen}{Innanen
  et~al.}{1997}]{Innanen97}
Innanen K.~A.,  Zheng J.~Q.,  Mikkola S.,    Valtonen M.~J.,  1997, AJ, 113,
  1915

\bibitem[\protect\citeauthoryear{{Kalas}, {Graham}, {Chiang}, {Fitzgerald},
  {Clampin}, {Kite}, {Stapelfeldt}, {Marois} \& {Krist}}{{Kalas}
  et~al.}{2008}]{Kalas08}
{Kalas} P.,  {Graham} J.~R.,  {Chiang} E.,  {Fitzgerald} M.~P.,  {Clampin} M.,
  {Kite} E.~S.,  {Stapelfeldt} K.,  {Marois} C.,    {Krist} J.,  2008, Science,
  322, 1345

\bibitem[\protect\citeauthoryear{King}{King}{1966}]{King66}
King I.~R.,  1966, AJ, 71, 64

\bibitem[\protect\citeauthoryear{Kouwenhoven, Brown, {Portegies Zwart} \&
  Kaper}{Kouwenhoven et~al.}{2007}]{Kouwenhoven07}
Kouwenhoven M. B.~N.,  Brown A. G.~A.,  {Portegies Zwart} S.~F.,    Kaper L.,
  2007, A\&A, 474, 77

\bibitem[\protect\citeauthoryear{Kouwenhoven, Brown, Zinnecker, Kaper \&
  {Portegies Zwart}}{Kouwenhoven et~al.}{2005}]{Kouwenhoven05}
Kouwenhoven M. B.~N.,  Brown A. G.~A.,  Zinnecker H.,  Kaper L.,    {Portegies
  Zwart} S.~F.,  2005, A\&A, 430, 137

\bibitem[\protect\citeauthoryear{Kozai}{Kozai}{1962}]{Kozai62}
Kozai Y.,  1962, AJ, 67, 591

\bibitem[\protect\citeauthoryear{{Kraus}, {Balega}, {Berger}, {Hofmann},
  {Millan-Gabet}, {Monnier}, {Ohnaka}, {Pedretti}, {Preibisch}, {Schertl},
  {Schloerb}, {Traub} \& {Weigelt}}{{Kraus} et~al.}{2007}]{Kraus07}
{Kraus} S.,  {Balega} Y.~Y.,  {Berger} J.-P.,  {Hofmann} K.-H.,  {Millan-Gabet}
  R.,  {Monnier} J.~D.,  {Ohnaka} K.,  {Pedretti} E.,  {Preibisch} T.,
  {Schertl} D.,  {Schloerb} F.~P.,  {Traub} W.~A.,    {Weigelt} G.,  2007,
  A\&A, 466, 649

\bibitem[\protect\citeauthoryear{{Kraus}, {Weigelt}, {Balega}, {Docobo},
  {Hofmann}, {Preibisch}, {Schertl}, {Tamazian}, {Driebe}, {Ohnaka}, {Petrov},
  {Sch{\"o}ller} \& {Smith}}{{Kraus} et~al.}{2009}]{Kraus09}
{Kraus} S.,  {Weigelt} G.,  {Balega} Y.~Y.,  {Docobo} J.~A.,  {Hofmann} K.-H.,
  {Preibisch} T.,  {Schertl} D.,  {Tamazian} V.~S.,  {Driebe} T.,  {Ohnaka} K.,
   {Petrov} R.,  {Sch{\"o}ller} M.,    {Smith} M.,  2009, A\&A, 497, 195

\bibitem[\protect\citeauthoryear{Kroupa}{Kroupa}{1995}]{Kroupa95a}
Kroupa P.,  1995, MNRAS, 277, 1491

\bibitem[\protect\citeauthoryear{Kroupa}{Kroupa}{2002}]{Kroupa02}
Kroupa P.,  2002, Science, 295, 82

\bibitem[\protect\citeauthoryear{Kroupa}{Kroupa}{2008}]{Kroupa08}
Kroupa P.,  2008, in {Aarseth} S.~J.,  {Tout} C.~A.,   {Mardling} R.~A.,  eds,
  Lecture Notes in Physics, Berlin Springer Verlag Vol.~760 of Lecture Notes in
  Physics, Berlin Springer Verlag, {Initial Conditions for Star Clusters}.
pp 181--259

\bibitem[\protect\citeauthoryear{Kroupa, Petr \& McCaughrean}{Kroupa
  et~al.}{1999}]{Kroupa99}
Kroupa P.,  Petr M.~G.,    McCaughrean M.~J.,  1999, New Astronomy, 4, 495

\bibitem[\protect\citeauthoryear{Lada}{Lada}{2010}]{Lada10}
Lada C.~J.,  2010, {Royal Society of London Philosophical Transactions Series
  A}, 368, 713

\bibitem[\protect\citeauthoryear{Lada \& Lada}{Lada \& Lada}{2003}]{Lada03}
Lada C.~J.,  Lada E.~A.,  2003, ARA\&A, 41, 57

\bibitem[\protect\citeauthoryear{{Lafreni{\`e}re}, {Doyon}, {Marois}, {Nadeau},
  {Oppenheimer}, {Roche}, {Rigaut}, {Graham}, {Jayawardhana}, {Johnstone},
  {Kalas}, {Macintosh} \& {Racine}}{{Lafreni{\`e}re}
  et~al.}{2007}]{Lafreniere07}
{Lafreni{\`e}re} D.,  {Doyon} R.,  {Marois} C.,  {Nadeau} D.,  {Oppenheimer}
  B.~R.,  {Roche} P.~F.,  {Rigaut} F.,  {Graham} J.~R.,  {Jayawardhana} R.,
  {Johnstone} D.,  {Kalas} P.~G.,  {Macintosh} B.,    {Racine} R.,  2007, ApJ,
  670, 1367

\bibitem[\protect\citeauthoryear{{Lafreni{\`e}re}, {Jayawardhana} \& {van
  Kerkwijk}}{{Lafreni{\`e}re} et~al.}{2010}]{Lafreniere10}
{Lafreni{\`e}re} D.,  {Jayawardhana} R.,    {van Kerkwijk} M.~H.,  2010, ApJ,
  719, 497

\bibitem[\protect\citeauthoryear{{Lagrange}, {Bonnefoy}, {Chauvin}, {Apai},
  {Ehrenreich}, {Boccaletti}, {Gratadour}, {Rouan}, {Mouillet}, {Lacour} \&
  {Kasper}}{{Lagrange} et~al.}{2010}]{Lagrange10}
{Lagrange} A.-M.,  {Bonnefoy} M.,  {Chauvin} G.,  {Apai} D.,  {Ehrenreich} D.,
  {Boccaletti} A.,  {Gratadour} D.,  {Rouan} D.,  {Mouillet} D.,  {Lacour} S.,
    {Kasper} M.,  2010, Science, 329, 57

\bibitem[\protect\citeauthoryear{Laughlin \& Adams}{Laughlin \&
  Adams}{1998}]{Laughlin98}
Laughlin G.,  Adams F.~C.,  1998, ApJ, 508, L171

\bibitem[\protect\citeauthoryear{{Lehmann}, {Vitrichenko}, {Bychkov},
  {Bychkova} \& {Klochkova}}{{Lehmann} et~al.}{2010}]{Lehmann10}
{Lehmann} H.,  {Vitrichenko} E.,  {Bychkov} V.,  {Bychkova} L.,    {Klochkova}
  V.,  2010, A\&A, 514, A34+

\bibitem[\protect\citeauthoryear{{Macintosh}, {Graham}, {Palmer}, {Doyon},
  {Gavel}, {Larkin}, {Oppenheimer}, {Saddlemyer}, {Wallace}, {Bauman}, {Evans},
  {Erikson}, {Morzinski} \& {Phillion}}{{Macintosh} et~al.}{2006}]{Macintosh06}
{Macintosh} B.,  {Graham} J.,  {Palmer} D.,  {Doyon} R.,  {Gavel} D.,  {Larkin}
  J.,  {Oppenheimer} B.,  {Saddlemyer} L.,  {Wallace} J.~K.,  {Bauman} B.,
  {Evans} J.,  {Erikson} D.,  {Morzinski} K.,    {Phillion} D. e.~a.,  2006, in
  Society of Photo-Optical Instrumentation Engineers (SPIE) Conference Series
  Vol.~6272 of Presented at the Society of Photo-Optical Instrumentation
  Engineers (SPIE) Conference, {The Gemini Planet Imager}

\bibitem[\protect\citeauthoryear{Malmberg, Davies \& Chambers}{Malmberg
  et~al.}{2007}]{Malmberg07a}
Malmberg D.,  Davies M.~B.,    Chambers J.~E.,  2007, MNRAS, 377, L1

\bibitem[\protect\citeauthoryear{{Marois}, {Macintosh}, {Barman}, {Zuckerman},
  {Song}, {Patience}, {Lafreni{\`e}re} \& {Doyon}}{{Marois}
  et~al.}{2008}]{Marois08}
{Marois} C.,  {Macintosh} B.,  {Barman} T.,  {Zuckerman} B.,  {Song} I.,
  {Patience} J.,  {Lafreni{\`e}re} D.,    {Doyon} R.,  2008, Science, 322, 1348

\bibitem[\protect\citeauthoryear{{Marois}, {Zuckerman}, {Konopacky},
  {Macintosh} \& {Barman}}{{Marois} et~al.}{2010}]{Marois10}
{Marois} C.,  {Zuckerman} B.,  {Konopacky} Q.~M.,  {Macintosh} B.,    {Barman}
  T.,  2010, Nature, 468, 1080

\bibitem[\protect\citeauthoryear{Mason, Gies, Hartkopf, W.~G.~Bagnuolo, {ten
  Brummelaar} \& McAlister}{Mason et~al.}{1998}]{Mason98}
Mason B.~D.,  Gies D.~R.,  Hartkopf W.~I.,  W.~G.~Bagnuolo J.,  {ten
  Brummelaar} T.,    McAlister H.~A.,  1998, AJ, 115, 821

\bibitem[\protect\citeauthoryear{Mason, Hartkopf, Gies, Henry \& Helsel}{Mason
  et~al.}{2009}]{Mason09}
Mason B.~D.,  Hartkopf W.~I.,  Gies D.~R.,  Henry T.~J.,    Helsel J.~W.,
  2009, AJ, 137, 3358

\bibitem[\protect\citeauthoryear{Mayor, Duquennoy, Halbwachs \&
  Mermilliod}{Mayor et~al.}{1992}]{Mayor92}
Mayor M.,  Duquennoy A.,  Halbwachs J.-L.,    Mermilliod J.-C.,  1992, in
  McAlister H.~A.,  Hartkopf W.~I.,  eds, {IAU Colloq. 135: Complementary
  Approaches to Double and Multiple Star Research} Vol.~32 of ASP Conference
  Series, {CORAVEL Surveys to Study Binaries of Different Masses and Ages}.
IAU, pp 73--81

\bibitem[\protect\citeauthoryear{Moraux, Lawson \& Clarke}{Moraux
  et~al.}{2007}]{Moraux07}
Moraux E.,  Lawson W.~A.,    Clarke C.~J.,  2007, A\&A, 473, 163

\bibitem[\protect\citeauthoryear{{Naoz}, {Farr}, {Lithwick}, {Rasio} \&
  {Teyssandier}}{{Naoz} et~al.}{2011}]{Naoz11}
{Naoz} S.,  {Farr} W.~M.,  {Lithwick} Y.,  {Rasio} F.~A.,    {Teyssandier} J.,
  2011, Nature, 473, 187

\bibitem[\protect\citeauthoryear{Parker \& Goodwin}{Parker \&
  Goodwin}{2009}]{Parker09c}
Parker R.~J.,  Goodwin S.~P.,  2009, MNRAS, 397, 1041

\bibitem[\protect\citeauthoryear{Parker, Goodwin \& Allison}{Parker
  et~al.}{2011}]{Parker11c}
Parker R.~J.,  Goodwin S.~P.,    Allison R.~J.,  2011, MNRAS, in press, arXiv:
  1108.3566

\bibitem[\protect\citeauthoryear{Parker, Goodwin, Kroupa \& Kouwenhoven}{Parker
  et~al.}{2009}]{Parker09a}
Parker R.~J.,  Goodwin S.~P.,  Kroupa P.,    Kouwenhoven M. B.~N.,  2009,
  MNRAS, 397, 1577

\bibitem[\protect\citeauthoryear{{Peretto}, {Andr{\'e}} \&
  {Belloche}}{{Peretto} et~al.}{2006}]{Peretto06}
{Peretto} N.,  {Andr{\'e}} P.,    {Belloche} A.,  2006, A\&A, 445, 979

\bibitem[\protect\citeauthoryear{Pfalzner \& Olczak}{Pfalzner \&
  Olczak}{2007}]{Pfalzner07}
Pfalzner S.,  Olczak C.,  2007, A\&A, 475, 875

\bibitem[\protect\citeauthoryear{Plummer}{Plummer}{1911}]{Plummer11}
Plummer H.~C.,  1911, MNRAS, 71, 460

\bibitem[\protect\citeauthoryear{{Portegies Zwart}, McMillan, Hut \&
  Makino}{{Portegies Zwart} et~al.}{2001}]{Zwart01}
{Portegies Zwart} S.~F.,  McMillan S. L.~W.,  Hut P.,    Makino J.,  2001,
  MNRAS, 321, 199

\bibitem[\protect\citeauthoryear{{Portegies Zwart}, Makino, McMillan \&
  Hut}{{Portegies Zwart} et~al.}{1999}]{Zwart99}
{Portegies Zwart} S.~F.,  Makino J.,  McMillan S. L.~W.,    Hut P.,  1999,
  A\&A, 348, 117

\bibitem[\protect\citeauthoryear{{Proszkow}, {Adams}, {Hartmann} \&
  {Tobin}}{{Proszkow} et~al.}{2009}]{Proszkow09}
{Proszkow} E.-M.,  {Adams} F.~C.,  {Hartmann} L.~W.,    {Tobin} J.~J.,  2009,
  ApJ, 697, 1020

\bibitem[\protect\citeauthoryear{{Quanz}, {Meyer}, {Kenworthy}, {Girard},
  {Kasper}, {Lagrange}, {Apai}, {Boccaletti}, {Bonnefoy}, {Chauvin}, {Hinz} \&
  {Lenzen}}{{Quanz} et~al.}{2010}]{Quanz10}
{Quanz} S.~P.,  {Meyer} M.~R.,  {Kenworthy} M.~A.,  {Girard} J.~H.~V.,
  {Kasper} M.,  {Lagrange} A.-M.,  {Apai} D.,  {Boccaletti} A.,  {Bonnefoy} M.,
   {Chauvin} G.,  {Hinz} P.~M.,    {Lenzen} R.,  2010, ApJ, 722, L49

\bibitem[\protect\citeauthoryear{Raghavan, McMaster, Henry, Latham, Marcy,
  Mason, Gies, White \& {ten Brummelaar}}{Raghavan et~al.}{2010}]{Raghavan10}
Raghavan D.,  McMaster H.~A.,  Henry T.~J.,  Latham D.~W.,  Marcy G.~W.,  Mason
  B.~D.,  Gies D.~R.,  White R.~J.,    {ten Brummelaar} T.~A.,  2010, ApJSS,
  190, 1

\bibitem[\protect\citeauthoryear{{Raymond}, {Armitage}, {Moro-Mart{\'{\i}}n},
  {Booth}, {Wyatt}, {Armstrong}, {Mandell}, {Selsis} \& {West}}{{Raymond}
  et~al.}{2011}]{Raymond11}
{Raymond} S.~N.,  {Armitage} P.~J.,  {Moro-Mart{\'{\i}}n} A.,  {Booth} M.,
  {Wyatt} M.~C.,  {Armstrong} J.~C.,  {Mandell} A.~M.,  {Selsis} F.,    {West}
  A.~A.,  2011, A\&A, 530, A62+

\bibitem[\protect\citeauthoryear{{Reggiani} \& {Meyer}}{{Reggiani} \&
  {Meyer}}{2011}]{Reggiani11}
{Reggiani} M.~M.,  {Meyer} M.~R.,  2011, ApJ, 738, 60

\bibitem[\protect\citeauthoryear{S{\'a}nchez \& Alfaro}{S{\'a}nchez \&
  Alfaro}{2009}]{Sanchez09}
S{\'a}nchez N.,  Alfaro E.~J.,  2009, ApJ, 696, 2086

\bibitem[\protect\citeauthoryear{Schmeja}{Schmeja}{2011}]{Schmeja11}
Schmeja S.,  2011, AN, 332, 172

\bibitem[\protect\citeauthoryear{Smith \& Bonnell}{Smith \&
  Bonnell}{2001}]{Smith01}
Smith K.~W.,  Bonnell I.~A.,  2001, MNRAS, 322, L1

\bibitem[\protect\citeauthoryear{Spurzem, Giersz, Heggie \& Lin}{Spurzem
  et~al.}{2009}]{Spurzem09}
Spurzem R.,  Giersz M.,  Heggie D.~C.,    Lin D. N.~C.,  2009, ApJ, 697, 458

\bibitem[\protect\citeauthoryear{Takeda \& Rasio}{Takeda \&
  Rasio}{2005}]{Takeda05}
Takeda G.,  Rasio F.~A.,  2005, ApJ, 627, 1001

\bibitem[\protect\citeauthoryear{Thies, Kroupa, Goodwin, Stamatellos \&
  Whitworth}{Thies et~al.}{2011}]{Thies11}
Thies I.,  Kroupa P.,  Goodwin S.~P.,  Stamatellos D.,    Whitworth A.~P.,
  2011, MNRAS, in press, arXiv: 1107.2113

\bibitem[\protect\citeauthoryear{{Weights}, {Lucas}, {Roche}, {Pinfield} \&
  {Riddick}}{{Weights} et~al.}{2009}]{Weights09}
{Weights} D.~J.,  {Lucas} P.~W.,  {Roche} P.~F.,  {Pinfield} D.~J.,
  {Riddick} F.,  2009, MNRAS, 392, 817

\end{thebibliography}

\label{lastpage}

\end{document}